%% file: main.tex
\documentclass[aip,amsmath,amssymb,reprint]{revtex4-1}

\usepackage{dcolumn}
\usepackage{bm}
\usepackage[mathlines]{lineno}
\usepackage[utf8]{inputenc}
\usepackage[T1]{fontenc}
\usepackage{mathptmx}
\usepackage{etoolbox}
\usepackage{graphicx}
\usepackage{booktabs}
\usepackage{amsmath}
\usepackage{amssymb}
\usepackage{braket}
\usepackage{stackengine}
\usepackage{subcaption}
\usepackage{xcolor}
\usepackage{xr}
\makeatletter
\def\@email#1#2{
	\endgroup
	\patchcmd{\titleblock@produce}
 	{\frontmatter@RRAPformat}
 	{\frontmatter@RRAPformat{\produce@RRAP{*#1\href{mailto:#2}{#2}}}\frontmatter@RRAPformat}
 	{}{}
}
\makeatother

\begin{document}
\input{molecules}

\title{A Length-Gauge Origin-Invariant Approach to Vibrational Circular Dichroism Spectra without Gauge-Including Atomic Orbitals}
\author{Brendan M. Shumberger}
	\affiliation{Department of Chemistry, Virginia Tech, Blacksburg, VA 24061, U.S.A.}
\author{James R. Cheeseman}
	\affiliation{Gaussian, Inc., 340 Quinnipiac Street, Building 40, Wallingford, CT 06492, U.S.A.}
\author{Marco Caricato}
	\affiliation{Department of Chemistry, University of Kansas, Lawrence, KS 66045, U.S.A.}
\author{T. Daniel Crawford}
	\affiliation{Department of Chemistry, Virginia Tech, Blacksburg, VA 24061, U.S.A.}
	\email{crawdad@vt.edu}
\date{\today}

\begin{abstract}
	\input{abstract}
\end{abstract}
\maketitle

\input{intro}
\input{theory}
\input{comp}
\input{results}
\input{conclusion}

\section{Supporting Information} \label{si}

Atomic coordinates, frequencies, dipole strengths, rotatory strengths, degrees of symmetry,
and VCD spectra of all test molecules

\section{Acknowledgements} \label{ack}

TDC was supported by the U.S.\ National Science Foundation via grant CHE-2154753
and BMS by grant OAC-2410880.
The authors are grateful to Advanced Research Computing at Virginia Tech for providing
computational resources that have contributed to the results reported within the paper.

\section{References}
\bibliography{references}

\end{document}

%% file: molecules.tex
\def\hhoo{(\textit{P})-hydrogen peroxide}
\def\smo{(\textit{S})-methyloxirane}
\def\apn{(1\textit{R}, 5\textit{R})-$\alpha$-pinene}
\def\cmp{(1\textit{R}, 4\textit{R})-camphor}

%% file: abstract.tex
We have extended the origin-invariant length gauge (LG(OI)) approach ---
originally developed by Caricato and co-workers for optical rotation (OR) and
electronic circular dichroism (ECD) --- to vibrational circular dichroism (VCD).
This approach avoids the need for gauge-including atomic orbitals (GIAOs), which
are typically required to circumvent the unphysical dependence of the CD rotatory
strengths on the arbitrary choice of coordinate origin for length gauge (LG)
computations.  Benchmark VCD spectra are presented for \hhoo, \smo, \apn, and
\cmp\ using Hartree-Fock (HF) theory and density functional theory (DFT) methods
across a range of basis sets and compared to those obtained from LG, velocity-gauge
(VG), and GIAO computations.  These analyses show that for VCD the LG(OI) approach
does not converge to the basis-set limit as rapidly as the GIAO approach, but does
yield similar quality spectra as GIAO for all major VCD peaks for
quadruple-zeta-quality basis sets.  The LG(OI) and VG VCD spectra are less
reliable compared to GIAOs for smaller basis sets.

%% file: intro.tex
\section{Introduction}

Vibrational circular dichroism (VCD) --- the differential absorption of left-
and right-circularly polarized infrared radiation --- is unique among
spectroscopies in that many of the experimental complications affecting the
discrimination of enatiomeric pairs are absent. Unlike X-ray crystallography,
VCD requires no high-quality single crystals; contrary to some chiral nuclear
magnetic resonance (NMR) methods, it requires no chiral derivatizing or
solvating agents; and in contrast to electronic circular dichroism (ECD), VCD
requires no UV-Vis chromophores, making it applicable to a broader range of
chiral molecules.\cite{Bogaerts2021, Polavarapu2016, Labuta2013}  As such, VCD
results are routinely accepted by the U.S. Food and Drug Administration (FDA)
as evidence confirming the absolute configuration of new drug
prospects.\cite{Wesolowski2013}  Due to the complex character of these results,
experimental evidence demands accompaniment by theoretical simulation.
\cite{Zhu2023}

Of primary concern in the simulation of VCD spectroscopy is the rotational
strength, which is the imaginary component of the dot product of the electric
and magnetic vibrational transition dipole moments.\cite{Stephens2000}  Given
the magnetic-field-dependent nature of this mixed dipole polarizability, one
must take care in formulating the rotational strength such that the molecular
property is invariant to shifts in the gauge origin.\cite{Ditchfield1974} In the
limit of a complete basis set, the rotational strength is invariant to the choice
of origin; however, for truncated basis sets, this is not the case.  For VCD,
the prevailing solution to this issue is the use of gauge-including atomic
orbitals (GIAOs) which are comprised of field dependent complex phase factors
multiplied onto the original atomic orbital basis functions, effectively
removing the unphysical origin dependence from the molecular Hamiltonian
integrals.\cite{Bak1993, Cheeseman1996, Helgaker1991}  Alternatively, the
distributed origin gauge with origins at nuclei provides much improved solutions
over those obtained by the common origin gauge, though these results still carry
some origin dependence.\cite{Stephens1987}

Recently, an origin invariant length-gauge (LG) approach, termed LG(OI), was
introduced by Caricato for optical rotation (OR)\cite{Caricato2020,
Caricato2021} and extended by Niemeyer, Caricato, and Neugebauer to ECD
\cite{Niemeyer2022}. These approaches yield origin invariance without the need
for GIAOs.  The premise behind the LG(OI) formulation is that one can combine
different orientations of the molecule in such a way that the mixed length
gauge/velocity gauge (LG/VG) dipole strength tensor, i.e.\ the component
containing the origin dependence when the origin is shifted, is diagonal,
thus nullifying any contribution from the asymmetry of the off-diagonal
elements initially contributing to the origin dependence of the rotational
strength calculated in the LG.\cite{Pelloni2014} However, rather than explicitly
rotating the molecule, one can perform a singular value decomposition (SVD) of
the mixed LG/VG dipole strength tensor to obtain the requisite rotational
matrices and subsequently apply them to the rotational strength tensor.

Until recently, simulations of VCD have been limited to Hartree-Fock (HF)
theory, multiconfigurational self-consistent field (MCSCF) theory, and density
functional theory (DFT) methods, all of which are formulated in the LG and rely
on the application of GIAOs to maintain origin invariance.\cite{Lowe1986,
Amos1987, Bak1993, Cheeseman1996}  To date, rotational strengths from the
velocity gauge representation of Stephens's formulation for VCD have not been
reported (at any level of theory) though the theoretical foundations have
already been laid out.\cite{Amos1988}  A formulation similar to Stephens's
magnetic field perturbation (MFP) theory,\cite{Stephens1985a} developed by
Nafie\cite{Nafie1992} and referred to as the nuclear velocity perturbation
theory (NVPT), was developed to specifically take advantage of the origin
invariance of the VG approach.  Its implementation and results for DFT have
recently been described by Ditler, Zimmermann, Kumar, and Luber\cite{Ditler2022}
and by Kumar and Luber.\cite{Kumar2025}  In Stephens's MFP formulation, the
electronic component of the electric-dipole transition moment in the VG, similar
to that of the magnetic-dipole transition moment, is formulated as an overlap
between wave function derivatives.  Unlike the magnetic-dipole transition
moment, however, the electric-dipole transition moment in this mixed derivative
overlap form requires that the derivative of the ket state wave function be
taken with respect to the magnetic vector potential.  It is this quantity which
is required to form the mixed LG/VG dipole strength tensor.

The purpose of this work is to extend Caricato's LG(OI) approach to the
simulation of VCD spectra in Stephens's formulation.  Given that the LG(OI)
method has not been utilized for VCD at any level of theory, we choose to focus
here on Hartree-Fock (HF) theory and density functional theory (DFT) methods so
that we may compare its performance to GIAO-based formulations. In the next
sections, we outline the theory of LG(OI) for VCD rotational strengths and then
examine results of spectral simulations using a range of basis sets for for test
compounds: \hhoo, \smo, \apn, and \cmp.

%% file: theory.tex
\section{Theory}
\subsection{Vibrational Circular Dichroism}
In simulations of VCD spectroscopy, the primary quantity of interest is the
rotational strength,
\begin{equation} \label{RotatoryStrength}
R_{Gg;Gk} = \textrm{Im} \left[ \braket{\Psi_{Gg} | \vec{\mu} | \Psi_{Gk}} \cdot \braket{\Psi_{Gk} | \vec{m} | \Psi_{Gg}} \right],
\end{equation}
which includes electric-dipole, $\braket{\Psi_{Gg} | \vec{\mu} | \Psi_{Gk}}$,
and magnetic-dipole, $\braket{\Psi_{Gk} | \vec{m} | \Psi_{Gg}}$, transition
moments between vibrational states $g$ and $k$ within the ground electronic
state, $G$.  In the vibrational harmonic approximation, the electric-dipole
transition moment of the $i$-th normal mode is given by the $\nu=0\rightarrow
1$ transition \cite{Wilson80}
\begin{equation} \label{ElectricDipoleTM}
    \braket{ 0 | \mu_{\beta} | 1 }_i = \left( \frac{\hbar}{2\omega_i} \right)^{1/2}
    \sum_{\lambda\alpha} P_{\alpha\beta}^{\lambda} S_{\lambda\alpha,i},
\end{equation}
where $\omega_i$ is the harmonic angular frequency associated with the normal
mode, $P_{\alpha\beta}^{\lambda}$ is the atomic polar tensor (APT), and
$S_{\lambda\alpha,i}$ is the normal coordinate transformation matrix from
Cartesian nuclear displacements to mass-weighted normal modes.  The subscripts
$\alpha$ and $\beta$ denote Cartesian directions of the nuclear coordinates and
external electric fields, respectively, while the superscript $\lambda$ indexes
the nuclei.  Using the electrical harmonic approximation, the APT is
\begin{equation} \label{APT}
    [P_{\alpha\beta}^\lambda]^{\text{LG}} = \left(\frac{\partial\braket{\Psi_G|\mu_{\beta}^e|\Psi_G}}{\partial R_{\lambda\alpha}}\right)_0 + Z_\lambda e \delta_{\alpha\beta},
\end{equation}
where $\braket{\Psi_G|\mu_{\beta}^e|\Psi_G}$ is the electronic component of the
electric-dipole moment expectation value, the subscript ``0'' denotes the
evaluation of the derivative at
the equilibrium/reference geometry, and $Z_\lambda e$ is the charge of the
$\lambda$-th nucleus.  To be explicit about the choice of gauge, we use the length
formulation of the electronic component of the electric dipole operator, i.e.
\begin{equation}
    \vec{\mu}^{e} = - e \sum_n \vec{r}_n,
\end{equation}
where $-e$ is the charge of the electron and $\vec{r}_n$ is the position
operator for the $n$-th electron.  Similarly, the magnetic-dipole transition
moment for the fundamental transition can be expressed as
\begin{equation} \label{MagneticDipoleTM}
    \braket{ 0 | m_{\beta} | 1 }_i = - \left( 2 \hbar^3 \omega_i \right)^{1/2}
    \sum_{\lambda\alpha} M_{\alpha\beta}^{\lambda} S_{\lambda\alpha,i},
\end{equation}
where $M_{\alpha\beta}^{\lambda}$ is the atomic axial tensor (AAT).  Indices
$\lambda$ and $\alpha$ have the same meaning as in Eq.~\eqref{APT}, however,
$\beta$ now denotes the Cartesian direction of the external magnetic field,
$\vec{H}$.  The AAT expands as\cite{Stephens1985}
\begin{align}
    M_{\alpha\beta}^{\lambda}
    & = \left\langle
    \left( \frac{\partial \Psi_G(\vec{R})}{\partial R_{\lambda\alpha}} \right)_0 \bigg|
    \left( \frac{\partial \Psi_G(\vec{R}_0, H_{\beta})}{\partial H_{\beta}} \right)_0 \right\rangle 
     \nonumber \\ & \quad + \sum_\gamma \epsilon_{\alpha\beta\gamma} R_{\lambda\gamma}^0 \frac{i Z_\lambda e}{4 \hbar}
\end{align}
where $\epsilon_{\alpha\beta\gamma}$ is the three-dimensional Levi-Civita
tensor, the subscripts ``0'' indicate that the derivatives are taken at the
equilibrium geometry and at zero magnetic field, and $R_{\lambda\gamma}^0$ is
the $\gamma$-th equilibrium Cartesian coordinate of the $\lambda$-th
nucleus.\cite{Stephens1985a}  In obtaining the electronic component of the
magnetic dipole transition moment, we use
\begin{equation}
    \vec{m}^e = - \frac{e}{2m} \sum_n \vec{r}_n \times \vec{p}_n
\end{equation}
as the form of the magnetic dipole operator.  Computation of the rotational
strength in an incomplete basis set using the LG APT without employing GIAOs
results in a quantity which depends on the choice of coordinate origin as
discussed in detail in a later section.

\subsection{Velocity-Gauge Electric-Dipole Transition Moment}

Following Amos, Jalkanen, and Stephens,\cite{Amos1987, Amos1988} the derivative of the expectation value of the electric-dipole operator may be written as
\begin{equation} \label{ElectricDipoleTM1}
    \left( \frac{\partial \left\langle \Psi_G | \mu_\beta^e | \Psi_G \right\rangle}{\partial R_{\lambda\alpha}} \right)_0
    = 2 \left\langle \left( \frac{\partial \Psi_G}{\partial R_{\lambda\alpha}} \right)_0 \bigg| \mu_\beta^e \bigg| \Psi_G^{(0)} \right\rangle,
\end{equation}
where the superscript $(0)$ on $\Psi_G^{(0)}$ denotes that the wave function
is evaluated at the equilibrium geometry and we have assumed real wave
functions.  A first-order perturbational expansion of the derivative of the wave function yields,
\begin{equation} \label{wfn nuc deriv}
    \left| \left( \frac{\partial \Psi_G}{\partial R_{\lambda\alpha}} \right)_{0} \right\rangle 
    = \sum_{K \neq G} \frac{\left\langle \Psi_K^{(0)} \left| \left( \frac{\partial H_{\text{el}}}{\partial R_{\lambda\alpha}} \right)_0 \right| \Psi_G^{(0)} \right\rangle }{W_G^{(0)} - W_K^{(0)}} \left| \Psi_K^{(0)} \right\rangle,
\end{equation}
and Eq.~\eqref{ElectricDipoleTM1} becomes
\begin{multline} \label{ElectricDipoleTM2}
    \left( \frac{\partial \left\langle \Psi_G | \mu_\beta^e | \Psi_G \right\rangle}{\partial R_{\lambda\alpha}} \right)_0
    = 2 \sum_{K \neq G} \frac{\left\langle \Psi_G^{(0)} \left| \left( \frac{\partial H_{\text{el}}}{\partial R_{\lambda\alpha}} \right)_0 \right| \Psi_K^{(0)} \right\rangle }{W_G^{(0)} - W_K^{(0)}} \\
    \times \left\langle \Psi_K^{(0)} \bigg| \mu_\beta^e \bigg| \Psi_G^{(0)} \right\rangle,
\end{multline}
where $H_{\text{el}}$ is the electronic Hamiltonian, and $W_G^{(0)}$ and $W_K^{(0)}$ are the electronic energies of the states $G$ and $K$, respectively.
We may use the off-diagonal hypervirial relation, 
\begin{equation}
    \left\langle \Psi_K^{(0)} | p_{n\beta} | \Psi_G^{(0)} \right\rangle = - \frac{i m}{\hbar} (W_G^{(0)} - W_K^{(0)}) \left\langle \Psi_K^{(0)} | r_{n\beta} | \Psi_G^{(0)} \right\rangle,
\end{equation}
to shift from the LG to the VG 
and Eq.~\eqref{ElectricDipoleTM2} becomes
\begin{align}
    \left( \frac{\partial \left\langle \Psi_G | \mu_\beta^e | \Psi_G \right\rangle}{\partial R_{\lambda\alpha}} \right)_0
    &= - \frac{2 i e \hbar}{m} \sum_{K \neq G} \frac{\left\langle \Psi_G^{(0)} \left| \left( \frac{\partial H_{\text{el}}}{\partial R_{\lambda\alpha}} \right)_0 \right| \Psi_K^{(0)} \right\rangle }{W_G^{(0)} - W_K^{(0)}} \nonumber \\
    &\quad \times \frac{\left\langle \Psi_K^{(0)} | \pi_\beta^e | \Psi_G^{(0)} \right\rangle }{W_G^{(0)} - W_K^{(0)}}
\end{align}
where we have defined the total electronic linear momentum operator as
\begin{equation}
    \vec{\pi}^e = \sum_n \vec{p}_{n}.
\end{equation}
Expanding the derivative of the electronic wave function with respect to the
vector potential, $\vec{A}$ (which enters the Hamiltonian through the spatially
uniform potential operator $\vec{A} \cdot \vec{\pi}^e$), to first order,
\begin{equation} \label{wfn elec deriv}
    \left| \left( \frac{\partial \Psi_G}{\partial A_\beta} \right)_{0} \right\rangle
      = \sum_{K \neq G} \frac{\left\langle \Psi_K^{(0)} | \pi_\beta^e | \Psi_G^{(0)} \right\rangle }{W_G^{(0)} - W_K^{(0)}} \left| \Psi_K^{(0)}
      \right\rangle,
\end{equation}
and adding the nuclear contribution yields the VG form of the APT,
\begin{equation}
    [P_{\alpha\gamma}^\lambda]^{\text{VG}}
    = - \frac{2 i e \hbar}{m} \left\langle \left( \frac{\partial \Psi_G}{\partial R_{\lambda\alpha}} \right)_{0} \bigg| \left( \frac{\partial \Psi_G}{\partial A_\gamma} \right)_{0} \right\rangle + Z_\lambda e \delta_{\alpha\gamma}.
\label{ElectricDipoleTM3}
\end{equation}
which is required for the LG(OI) approach and VG approach using Stephens's
definition of the AAT, as shown in the next section.

\subsection{The LG(OI) Approach for the Vibrational Rotatory Strength}

While both the LG and VG APTs --- $[P_{\alpha\gamma}^\lambda]^{\text{LG}}$ and
$[P_{\alpha\gamma}^\lambda]^{\text{VG}}$, respectively --- are origin-independent,
the AAT is not.  As shown by Stephens\cite{Stephens1987} and by Amos, Jalkanen,
and Stephens,\cite{Amos1987} a shift of the coordinate origin at $\vec{O}_1$
along a vector $\vec{B}$ to a new origin, $\vec{O}_2$, yields a shift in the
AAT,
\begin{equation} \label{M4}
    (M_{\alpha\beta}^\lambda)^{O_2}
    = (M_{\alpha\beta}^\lambda)^{O_1}
      + \frac{i}{4 \hbar} \sum_{\sigma\gamma} \epsilon_{\sigma\beta\gamma} B_\gamma [P_{\alpha\sigma}^\lambda]^{\text{VG}},
\end{equation}
affecting the rotational strength of the $i$-th normal mode,
\begin{equation} \label{rot str}
    R_i = \text{Im} \left[ \left\langle 0 | \vec{\mu} | 1 \right\rangle_i \cdot \left\langle 1 | \vec{m} | 0 \right\rangle_i \right].
\end{equation}
as
\begin{equation} \label{LGVG RotStr}
    [R_i^{O_2}]^{\text{LG}}
    = [R_i^{O_1}]^{\text{LG}}
      + \frac{\hbar}{4} \vec{B} \cdot [\vec{P}_i]^{\text{LG}} \times [\vec{P}_i]^{\text{VG}},
\end{equation}
where the superscript LG on $[R_i^{O_2}]^{\text{LG}}$ indicates the gauge chosen
for the APT component of the rotational strength.  Thus, the LG rotational strength 
is not origin invariant for truncated basis sets because $[\vec{P}_i]^{\text{LG}}
\neq [\vec{P}_i]^{\text{VG}}$.  However, if one chooses to compute the rotational
strength using the VG for the APT, the rotational strength is invariant to changes
in the coordinate origin:
\begin{equation}
    [R_i^{O_2}]^{\text{VG}}
    = [R_i^{O_1}]^{\text{VG}}
      + \frac{\hbar}{4} \vec{B} \cdot [\vec{P}_i]^{\text{VG}} \times [\vec{P}_i]^{\text{VG}}
      = [R_i^{O_1}]^{\text{VG}}.
\end{equation}

In the LG(OI) approach, one can remove the origin dependence of the LG rotational
strength on the second term by diagonalizing the elements of the matrix formed
from the outer product of $[\vec{P}_i]^{\text{LG}}$ and $[\vec{P}_i]^{\text{VG}}$, i.e.
\begin{equation} \label{SVD}
    \mathbf{[D_i]^{\text{LG/VG}}} = [\vec{P}_i]^{\text{LG}} \otimes [\vec{P}_i]^{\text{VG}}.
\end{equation}
This may be viewed as effectively reorienting the molecule along the principal
axis of the mixed LG/VG tensor such that the second term
in Eq.~\eqref{LGVG RotStr} is zero, rendering the approach origin invariant.
The diagonalization is performed by an SVD from
which one obtains the diagonalized matrix, $\mathbf{[D_i']^{\text{LG/VG}}}$, and the
unitary transformation matrices, $\mathbf{U_i}$ and $\mathbf{V_i^\dagger}$ such that
\begin{equation} \label{SVD1}
    \mathbf{[D_i']^{\text{LG/VG}} = U_i^\dagger [D_i]^{\text{LG/VG}} V_i}.
\end{equation}
Noting that $[R_i]^{\text{LG}} = \text{tr}(\mathbf{[R_i]^{\text{LG}}})$ we must
perform the same unitary transformation on $\mathbf{[R_i]^{\text{LG}}}$
to obtain the LG(OI) rotational strength for the $i$-th normal mode,
\begin{equation}
    [R_i']^{\text{LG(OI)}} = \text{tr}(\mathbf{U_i^\dagger [R_i]^{\text{LG}} V_i}),
\end{equation}
which is origin invariant.

%% file: comp.tex
\section{Computational Details}
\begin{figure*}[!hbtp]
    \centering
    \includegraphics[width=6.75in,height=2.15in]{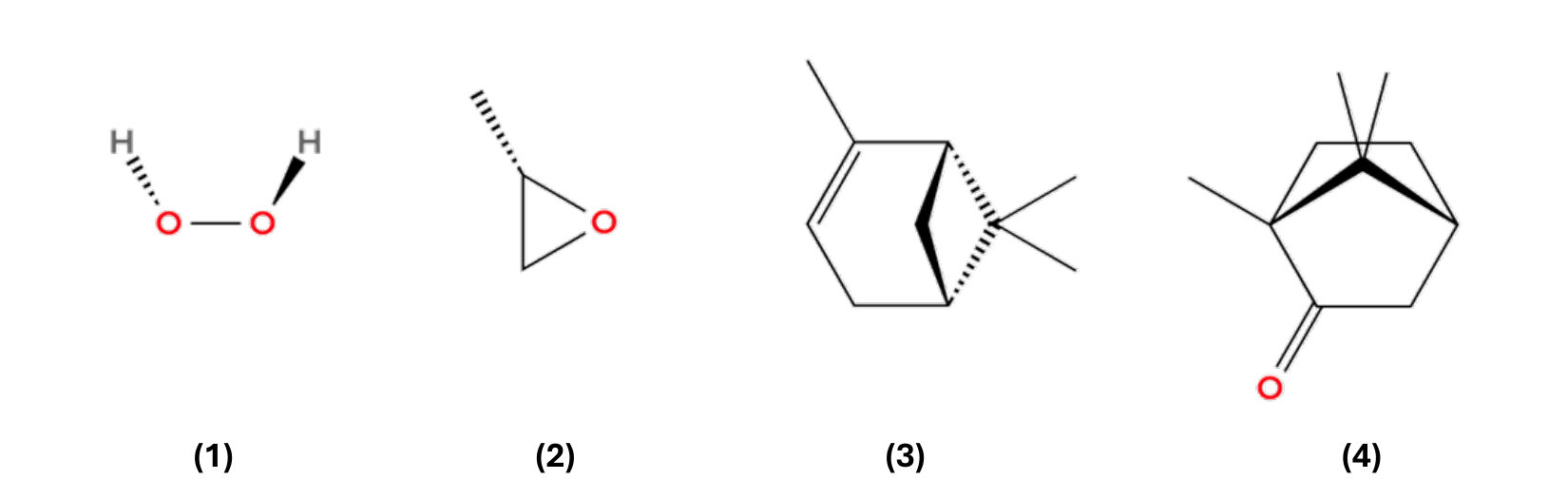}
    \caption{Molecular test set for simulation of LG-, VG-, LG(OI)-, and GIAO-based
    simulations of VCD spectra including \textbf{(1)} \hhoo, \textbf{(2)} \smo, 
    \textbf{(3)} \apn, and \textbf{(4)} \cmp.}
    \label{fig:molecular_test_set}
\end{figure*}

In this work, we compare the rotational strengths computed using the LG(OI) 
approach with those from the LG-, VG-, and GIAO-based approaches.  Our molecular
test set, as shown in Fig.~\ref{fig:molecular_test_set} includes \hhoo, \smo,
\apn, and \cmp.  Due to their inherent rigidity, historically \smo, \apn, and
\cmp experimental spectra have been used to benchmark HF and DFT simulations of
VCD spectra.\cite{Stephens2012}  We continue in this vein, though we use
simulated GIAO-based spectra to benchmark the HF and DFT implementations of the
VG and LG(OI) approaches.  Our analysis of these three molecules involves
calculations using the HF and DFT (B3PW91) methods using the cc-pVTZ,
aug-cc-pVTZ, and aug-cc-pVQZ basis sets.\cite{Dunning1989, Kendall1992a}
Additionally, we include HF calculations of \hhoo\ since it allows us to show
convergence of the rotational strengths for the various methods across a range of basis sets, 
including aug-cc-pVDZ, aug-cc-pVTZ, aug-cc-pVQZ, aug-cc-pV5Z,
aug-cc-pV6Z, UGBS1P+2+, and UGBS2P+2+.\cite{Dunning1989,
Kendall1992a, Peterson1994a, Wilson1996a, Mourik1999a, Castro1998a}  

A potential measure of the convergence of the VG and LG(OI) methods towards the
basis set limit is the degree of symmetry (DoS) of the $[D_i]_A^{\text{LG/VG}}$
matrix, which is based on that defined by Caricato and
Balduf,\cite{Caricato2021} \textit{viz.}
\begin{equation}
\Delta_i^s = 1 - \frac{\left\lVert[D_i]_A^{\text{LG/VG}}\right\rVert_F}{\left\lVert[D_i]^{\text{LG/VG}}\right\rVert_F}.
\end{equation}
where $\left\lVert\cdot\right\rVert_F$ represents the Frobenius norm of the
given matrix and the subscript $A$ denotes the antisymmetric part of the
$[D_i]^{\text{LG/VG}}$ matrix.  Ideally, as the basis set is expanded towards
completeness, $[D_i]_A^{\text{LG/VG}}$ should become more and more symmetric,
and thus the value of $\Delta_i^s$ for each mode should approach unity. 

All geometry optimizations and frequency, dipole strength, and rotational
strength calculations were performed using a consistent level of theory/basis
set.  Likewise, these geometries, frequencies, dipole strengths, rotational
strengths, and spectra for all calculations are included in the supporting
information.  All simulations were performed with a development version of
the GAUSSIAN suite of programs.\cite{GaussianDevVer}  The LG(OI) transformations
were executed with external scripts.

%% file: results.tex
\section{Results}

\subsection{Origin Invariance}

\begin{figure*}
\centering
    \begin{subfigure}[t]{\textwidth}
	\stackunder[3pt]{\includegraphics[width=6.4in,height=2.5in]{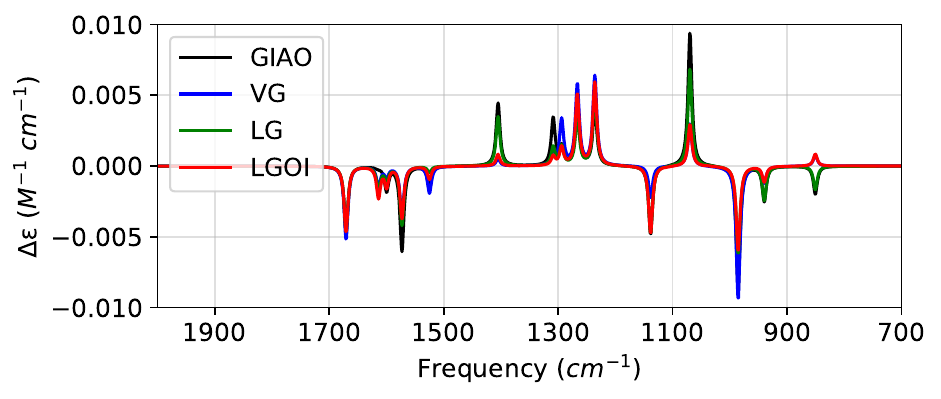}}{\
	\ \ \ \ {\large (a)}}
    \end{subfigure} \\ \begin{subfigure}[t]{\textwidth}
	\stackunder[3pt]{\includegraphics[width=6.4in,height=2.5in]{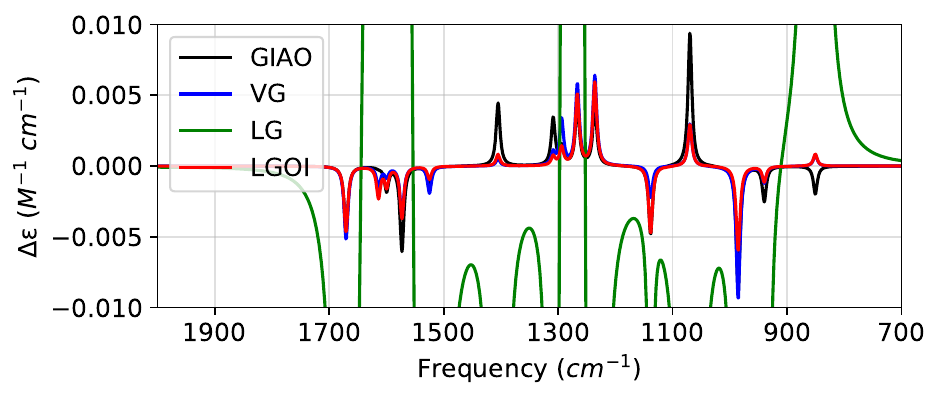}}{\
	\ \ \ \ {\large (b)}}
    \end{subfigure}
\caption{VCD spectra of \smo\ optimized and computed with the HF method using
the
	aug-cc-pVTZ basis set for geometries with the coordinate origin located at (a) the center of mass
	and (b) translated by 1000 a.u.\ in each Cartesian direction.}
\label{fig:smo_origin_invariance_spectra} \end{figure*}

In Fig.~\ref{fig:smo_origin_invariance_spectra} and
Table~\ref{tab:smo_hf_rot_str} we demonstrate the origin invariance of the
LG(OI) method as applied to VCD using the \smo\ test case and the aug-cc-pVTZ
basis set.  As expected, LG rotational strengths change drastically when the
gauge origin is shifted away from the molecule's center of mass.  The VG and
LG(OI) formulations of the rotational strength are found to exhibit exact origin
invariance while the GIAO method yields only slight origin dependence which is
attributed to numerical precision.  We note that there is only one sign
discrepancy in the data presented in
Fig.~\ref{fig:smo_origin_invariance_spectra} between the LG(OI) and VG methods
vs.\ the GIAO method for mode four which primarily involves stretching motions
of the C$-$C bond between the methyl group and epoxide ring and the C$-$O bond
between the oxygen and primary carbon in the epoxide ring.

\begin{table*}
    \scriptsize
    \caption{Frequencies (cm$^{-1}$) and rotational strengths (10$^{-44}$ esu$^{2}$
            cm$^{2}$) of \smo\ computed with the HF method using the
            aug-cc-pVTZ basis set with the coordinate origin located at the center of mass 
            ($R_i^{O_1}$) compared to that with the origin translated by 1000 a.u.\ in each Cartesian direction
            ($R_i^{O_2}$).}
    \label{tab:smo_hf_rot_str}
    \renewcommand{\arraystretch}{0.6}
    \begin{tabular*}{\textwidth}{@{\extracolsep{\fill}}lrrrrrrrrr}
    \toprule
          & & \multicolumn{2}{c}{LG} & \multicolumn{2}{c}{VG} & \multicolumn{2}{c}{GIAO} & \multicolumn{2}{c}{LGOI} \\
    \cmidrule{3-4} \cmidrule{5-6} \cmidrule{7-8} \cmidrule{9-10}
    Mode  & \multicolumn{1}{c}{Frequency} & \multicolumn{1}{c}{$R_i^{O_1}$} & \multicolumn{1}{c}{$R_i^{O_2}$} & \multicolumn{1}{c}{$R_i^{O_1}$} & \multicolumn{1}{c}{$R_i^{O_2}$} & \multicolumn{1}{c}{$R_i^{O_1}$} & \multicolumn{1}{c}{$R_i^{O_2}$} & \multicolumn{1}{c}{$R_i^{O_1}$} & \multicolumn{1}{c}{$R_i^{O_2}$} \\
    \midrule
     1 &  227.25 &  -3.550 &   -92.732 &  -5.251 &  -5.251 &  -3.488 &  -3.490 &  -5.028 &  -5.028  \\
     2 &  397.04 &  13.246 &  1719.289 &  15.173 &  15.173 &  12.651 &  12.649 &  19.039 &  19.039  \\
     3 &  441.78 &   8.071 &  -269.916 &   3.752 &   3.752 &   7.372 &   7.373 &   3.882 &   3.882  \\
     4 &  850.65 &  -5.670 &  2711.164 &   2.702 &   2.702 &  -6.749 &  -6.741 &   2.913 &   2.913  \\
     5 &  939.82 &  -7.401 &  -474.200 &  -3.581 &  -3.581 &  -7.616 &  -7.611 &  -3.402 &  -3.402  \\
     6 &  985.48 & -18.203 &  -591.866 & -27.668 & -27.668 & -25.968 & -25.972 & -17.613 & -17.613  \\
     7 & 1069.72 &  18.619 & -1928.606 &   8.132 &   8.132 &  25.522 &  25.522 &   8.010 &   8.010  \\
     8 & 1138.48 & -12.092 &   -74.257 &  -5.749 &  -5.749 & -12.371 & -12.375 & -12.113 & -12.113  \\
     9 & 1235.94 &  11.888 & -1737.372 &  14.757 &  14.757 &  10.006 &  10.014 &  13.713 &  13.713  \\
    10 & 1266.46 &   7.994 &  1250.630 &  13.030 &  13.030 &  10.040 &  10.039 &  11.464 &  11.464  \\
    11 & 1293.93 &   3.005 &    83.988 &   7.165 &   7.165 &   2.805 &   2.805 &   2.836 &   2.836  \\
    12 & 1308.67 &   2.875 &  -534.910 &   1.917 &   1.917 &   7.388 &   7.387 &   1.421 &   1.421  \\
    13 & 1405.00 &   7.188 &  -833.769 &   1.067 &   1.067 &   9.146 &   9.147 &   1.714 &   1.714  \\
    14 & 1524.87 &  -0.803 & -3491.393 &  -3.611 &  -3.611 &  -3.049 &  -3.053 &  -1.738 &  -1.738  \\
    15 & 1572.97 &  -7.657 &  1107.651 &  -6.294 &  -6.294 & -11.002 & -10.997 &  -6.730 &  -6.730  \\
    16 & 1599.76 &  -1.707 &  2176.053 &  -1.291 &  -1.291 &  -3.094 &  -3.092 &  -2.470 &  -2.470  \\
    17 & 1613.92 &  -2.908 &  1611.164 &  -2.496 &  -2.496 &  -0.138 &  -0.143 &  -3.888 &  -3.888  \\
    18 & 1670.61 &  -7.957 & -1861.923 &  -8.951 &  -8.951 &  -8.089 &  -8.089 &  -8.036 &  -8.036  \\
    19 & 3161.71 &  -1.158 &  -582.917 &  -1.569 &  -1.569 &  -1.117 &  -1.107 &  -1.270 &  -1.270  \\
    20 & 3217.28 &   7.840 &   402.944 &  10.032 &  10.032 &   7.218 &   7.223 &   7.260 &   7.260  \\
    21 & 3231.19 & -13.371 &  -103.315 & -18.970 & -18.970 & -11.259 & -11.255 & -12.221 & -12.221  \\
    22 & 3242.37 &  19.453 &  1552.329 &  24.879 &  24.880 &  14.979 &  14.934 &  19.742 &  19.742  \\
    23 & 3261.55 & -14.356 &  -463.911 & -16.017 & -16.017 & -10.482 & -10.477 & -12.171 & -12.171  \\
    24 & 3325.42 &   6.248 & -1037.726 &   8.311 &   8.311 &   5.502 &   5.529 &   5.822 &   5.822  \\
    \bottomrule
    \end{tabular*}
\end{table*}

\subsection{Basis Set Convergence}

\begin{figure*}
\centering
    \begin{subfigure}[t]{0.30\textwidth}
        \stackunder[3pt]{\includegraphics[width=2in,height=2in]{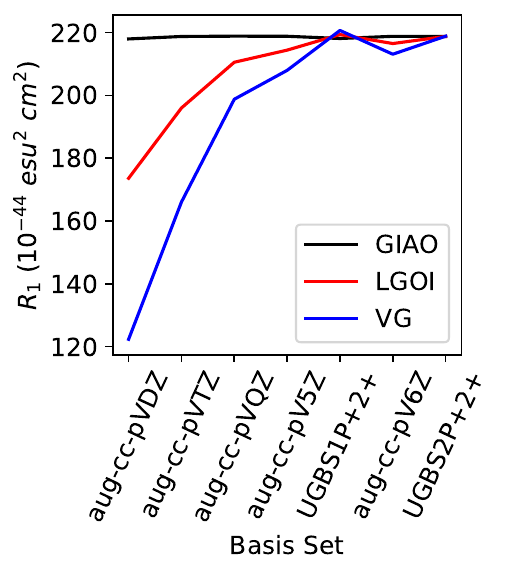}}{\ \ \ \ \ {\large (a)}}
    \end{subfigure}
    \begin{subfigure}[t]{0.30\textwidth}
        \stackunder[3pt]{\includegraphics[width=2in,height=2in]{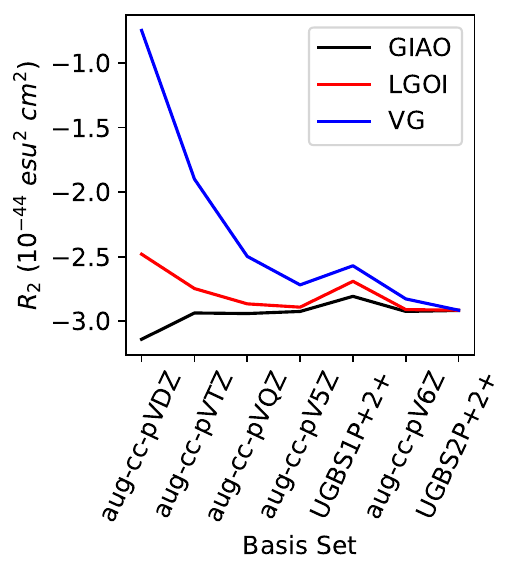}}{\ \ \ \ \ {\large (b)}}
    \end{subfigure} 
    \begin{subfigure}[t]{0.30\textwidth}
        \stackunder[3pt]{\includegraphics[width=2in,height=2in]{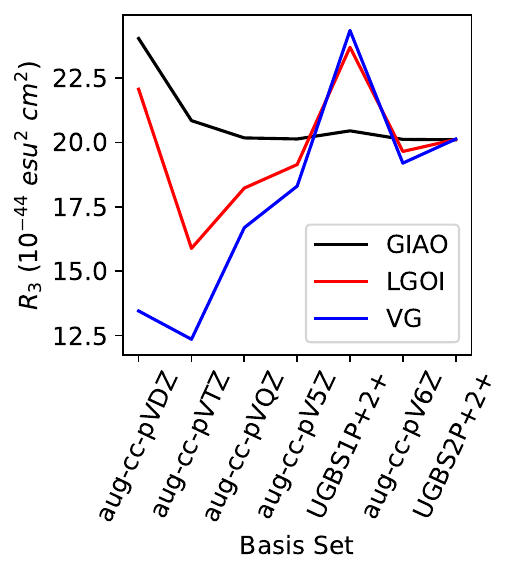}}{\ \ \ \ \ {\large (c)}}
    \end{subfigure} \\
    \begin{subfigure}[t]{0.30\textwidth}
        \stackunder[3pt]{\includegraphics[width=2in,height=2in]{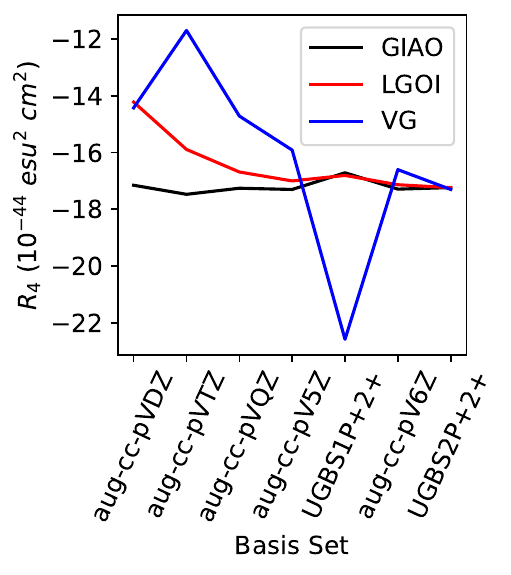}}{\ \ \ \ \ {\large (d)}}
    \end{subfigure}
    \begin{subfigure}[t]{0.30\textwidth}
        \stackunder[3pt]{\includegraphics[width=2in,height=2in]{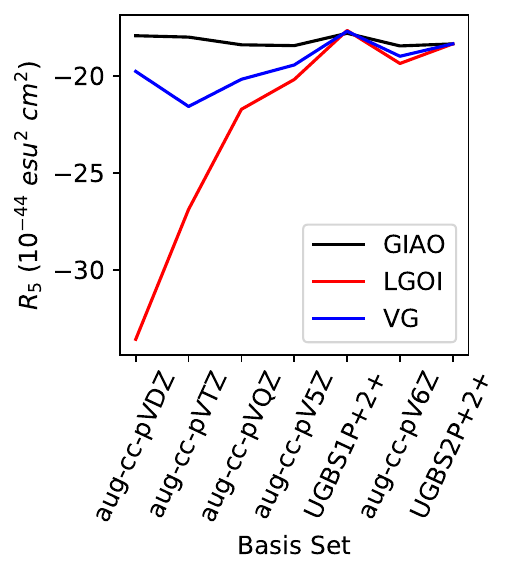}}{\ \ \ \ \ {\large (e)}}
    \end{subfigure} 
    \begin{subfigure}[t]{0.30\textwidth}
        \stackunder[3pt]{\includegraphics[width=2in,height=2in]{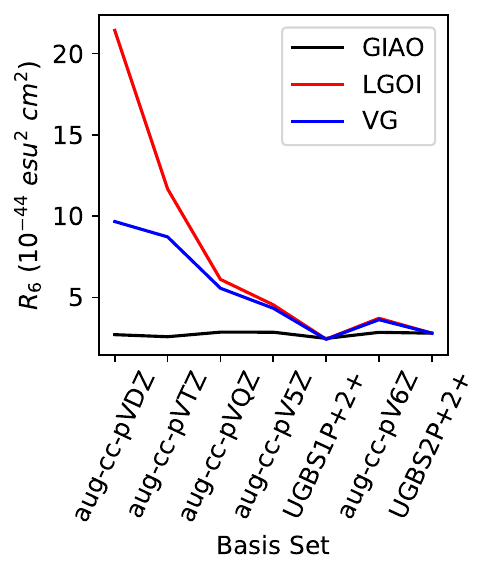}}{\ \ \ \ \ {\large (f)}}
    \end{subfigure}
\caption{Basis-set convergence of the rotational strength of each normal mode of
\hhoo: (a) H$-$O$-$O$-$H torsion, (b) O$-$O stretch, (c) H$-$O$-$O
antisymmetric bend, (d) H$-$O$-$O symmetric bend, (e) O$-$H antisymmetric
stretch, and (f) O$-$H symmetric stretch.}

\label{fig:hhoo_hf_convergence}
\end{figure*}

Vibrational frequencies, dipole strengths, rotational strengths, and degrees of
symmetry for \hhoo\ are given in Table~\ref{tab:hhoo_hf_rot_str} for the aug-cc-pVDZ,
aug-cc-pVTZ, aug-cc-pVQZ, aug-cc-pV5Z, aug-cc-pV6Z, UGBS1P+2+, and UGBS2P+2+
basis sets.  In addition, we include basis-set convergence behavior for each
of the six modes of this molecule in Fig.~\ref{fig:hhoo_hf_convergence}.  We
observe clear convergence towards the basis set limit for all three origin
invariant methods including the VG, LG(OI), and GIAO based approaches.
However, the GIAO method converges much more rapidly than both the VG and
LG(OI) approaches.  The LG(OI) approach provides a compromising convergence
pattern between the VG approach and the GIAO-based approach in four of the six
modes.  Interestingly, the out-of-phase and in-phase hydrogen stretching vibrations 
(modes five and six), display faster convergence behavior for the VG
approach over that of LG(OI).

Of note are the degrees of symmetry for which four of the six modes had
values of exactly $1.000$ indicating that the orientation of the molecule was
such that the mixed dipole strength tensor was already symmetric --- in this
case diagonal, as is evident from the fact that the LG and LG(OI) rotational
strengths are identical for these modes.  Unfortunately, the degree of
symmetry does not provide a very sensitive measure of convergence of the
rotational strengths even for the other two vibrational modes of \hhoo, and it
even decreases for mode three (H$-$O$-$O antisymmetric bending) between the
aug-cc-pVDZ and aug-cc-pVTZ basis sets.  However, for larger molecules with
more vibrational degrees of freedom it is significantly more effective, as
discussed below.

Given the remaining disparity between rotational strengths between the VG,
LG(OI), and GIAO approaches for these modes, one can ascribe the differences
to the ability of the basis set to describe the space on which the
different operators act in the different formulations of the rotational
strength.  In the LG formulation, one evaluates integrals associated with the
position and angular momentum operators while the VG formalism requires
evaluation of the linear momentum operator instead of the position operator.
The LG(OI) approach requires evaluation of all these integrals.  In contrast,
the GIAO approach evaluates modified origin invariant integrals.  From
Fig.~\ref{fig:hhoo_hf_convergence}, it is clear that in the GIAO formulation
provides a better description of the space on which the operators acts.


\begin{table*}
    \scriptsize
    \caption{Frequencies (cm$^{-1}$), dipole strengths (10$^{-40}$ esu$^{2}$ cm$^{2}$), rotational
        strengths (10$^{-44}$ esu$^{2}$ cm$^{2}$), and degrees of symmetry for \hhoo.}
    \label{tab:hhoo_hf_rot_str}
    \renewcommand{\arraystretch}{0.6}
    \begin{tabular*}{\textwidth}{@{\extracolsep{\fill}}lrrrrrrrrrrr}
    \toprule
          & &           & \multicolumn{3}{c}{Dipole Strength} & \hspace{1pt} & \multicolumn{4}{c}{Rotatory Strength} \\
    \cmidrule{4-6} \cmidrule{8-11}
    Mode  & \multicolumn{1}{c}{Frequency} & & \multicolumn{1}{c}{LG} & \multicolumn{1}{c}{VG} & \multicolumn{1}{c}{Mixed} & & \multicolumn{1}{c}{LG} & \multicolumn{1}{c}{VG} & \multicolumn{1}{c}{GIAO} & \multicolumn{1}{c}{LGOI} & \multicolumn{1}{c}{DoS} \\ \midrule
    \multicolumn{12}{c}{HF/aug-cc-pVDZ} \\
    \cmidrule{0-11}
    1 &  423.60 & & 1826.696 & 906.888 & 1287.093 & & 173.595 & 122.315 & 217.985 & 173.595 &   1.000 \\
    2 & 1139.88 & &   2.886  &   0.262 &   0.869  & &  -2.481 &  -0.747 &  -3.140 &  -2.481 &   1.000 \\
    3 & 1491.09 & & 282.332  & 104.976 & 172.151  & &  20.645 &  13.456 &  24.037 &  22.067 &   0.994 \\
    4 & 1608.11 & &   0.978  &   1.006 &   0.992  & & -14.220 & -14.424 & -17.153 & -14.220 &   1.000 \\
    5 & 4139.34 & &  91.145  &  31.536 &  52.657  & & -38.579 & -19.746 & -17.905 & -33.569 &   0.867 \\
    6 & 4139.72 & &  26.902  &   5.482 &  12.144  & &  21.424 &   9.671 &   2.713 &  21.424 &   1.000 \\
    \cmidrule{0-11}
    \multicolumn{12}{c}{HF/aug-cc-pVTZ} \\
    \cmidrule{0-11}
    1 &  411.47 & & 1849.172 & 1327.068 & 1566.517 & & 195.996 & 166.037 & 218.770 & 195.996 &   1.000 \\
    2 & 1159.85 & &   2.535  &   1.210  &   1.752  & &  -2.747 &  -1.898 &  -2.936 &  -2.747 &   1.000 \\
    3 & 1487.66 & & 278.948  & 168.704  & 216.894  & &  19.457 &  12.355 &  20.848 &  15.887 &   0.987 \\
    4 & 1604.18 & &   1.014  &   0.549  &   0.747  & & -15.891 & -11.695 & -17.473 & -15.891 &   1.000 \\
    5 & 4127.68 & &  92.361  &  59.455  &  73.978  & & -27.881 & -21.555 & -17.977 & -26.866 &   0.959 \\
    6 & 4128.46 & &  26.878  &  15.014  &  20.089  & &  11.674 &   8.725 &   2.586 &  11.674 &   1.000 \\
    \cmidrule{0-11}
    \multicolumn{12}{c}{HF/aug-cc-pVQZ} \\
    \cmidrule{0-11}
    1 &  411.98 & & 1841.504 & 1641.183 & 1738.460 & & 210.561 & 198.779 & 218.871 & 210.561 &   1.000 \\
    2 & 1162.23 & &   2.550  &   1.937  &   2.222  & &  -2.866 &  -2.498 &  -2.941 &  -2.866 &   1.000 \\
    3 & 1491.70 & & 277.808  & 232.930  & 254.375  & &  19.675 &  16.690 &  20.177 &  18.227 &   0.995 \\
    4 & 1608.34 & &   0.989  &   0.769  &   0.872  & & -16.685 & -14.713 & -17.258 & -16.685 &   1.000 \\
    5 & 4135.12 & &  92.073  &  79.405  &  85.492  & & -21.923 & -20.153 & -18.374 & -21.701 &   0.988 \\
    6 & 4135.96 & &  26.868  &  22.288  &  24.471  & &   6.113 &   5.568 &   2.869 &   6.113 &   1.000 \\
    \cmidrule{0-11}
    \multicolumn{12}{c}{HF/aug-cc-pV5Z} \\
    \cmidrule{0-11}
    1 &  411.15 & & 1848.646 & 1739.227 & 1793.102 & & 214.421 & 207.979 & 218.837 & 214.421 &   1.000 \\
    2 & 1162.72 & &   2.535  &   2.241  &   2.384  & &  -2.891 &  -2.718 &  -2.924 &  -2.891 &   1.000 \\
    3 & 1492.35 & & 277.962  & 254.169  & 265.798  & &  19.883 &  18.304 &  20.135 &  19.141 &   0.997 \\
    4 & 1609.43 & &   0.993  &   0.870  &   0.929  & & -17.001 & -15.914 & -17.302 & -17.001 &   1.000 \\
    5 & 4136.95 & &  92.104  &  85.376  &  88.673  & & -20.261 & -19.413 & -18.416 & -20.163 &   0.994 \\
    6 & 4137.72 & &  26.752  &  24.329  &  25.512  & &   4.548 &   4.337 &   2.864 &   4.548 &   1.000 \\
    \cmidrule{0-11}
    \multicolumn{12}{c}{HF/aug-cc-pV6Z} \\
    \cmidrule{0-11}
    1 &  410.99 & & 1850.140 & 1792.430 & 1821.056 & & 216.518 & 213.115 & 218.822 & 216.518 &   1.000 \\
    2 & 1162.92 & &   2.533  &   2.391  &   2.461  & &  -2.910 &  -2.827 &  -2.924 &  -2.910 &   1.000 \\
    3 & 1492.44 & & 278.038  & 265.384  & 271.637  & &  19.993 &  19.198 &  20.120 &  19.650 &   0.999 \\
    4 & 1609.65 & &   0.991  &   0.931  &   0.960  & & -17.135 & -16.602 & -17.290 & -17.135 &   1.000 \\
    5 & 4137.03 & &  92.132  &  88.623  &  90.360  & & -19.381 & -18.965 & -18.432 & -19.337 &   0.997 \\
    6 & 4137.78 & &  26.739  &  25.482  &  26.103  & &   3.721 &   3.633 &   2.855 &   3.721 &   1.000 \\
    \cmidrule{0-11}
    \multicolumn{12}{c}{HF/UGBS1P+2+} \\
    \cmidrule{0-11}
    1 &  429.13 & & 1760.491 & 1781.663 & 1771.046 & & 219.377 & 220.692 & 218.125 & 219.377 &   1.000 \\
    2 & 1156.56 & &   2.446  &   2.232  &   2.336  & &  -2.691 &  -2.570 &  -2.807 &  -2.691 &   1.000 \\
    3 & 1496.69 & & 277.755  & 293.283  & 285.376  & &  20.331 &  24.351 &  20.452 &  23.697 &   0.989 \\
    4 & 1610.77 & &   0.923  &   1.666  &   1.240  & & -16.807 & -22.585 & -16.714 & -16.807 &   1.000 \\
    5 & 4134.37 & &  92.304  &  92.739  &  92.519  & & -17.700 & -17.674 & -17.779 & -17.632 &   0.995 \\
    6 & 4135.52 & &  27.603  &  27.394  &  27.499  & &   2.437 &   2.427 &   2.482 &   2.437 &   1.000 \\
    \cmidrule{0-11}
    \multicolumn{12}{c}{HF/UGBS2P+2+} \\
    \cmidrule{0-11}
    1 &  411.45 & & 1846.562 & 1847.672 & 1847.117 & & 218.798 & 218.864 & 218.749 & 218.798 &   1.000 \\
    2 & 1162.28 & &   2.524  &   2.524  &   2.524  & &  -2.915 &  -2.915 &  -2.917 &  -2.915 &   1.000 \\
    3 & 1492.23 & & 277.916  & 278.278  & 278.097  & &  20.103 &  20.130 &  20.103 &  20.117 &   1.000 \\
    4 & 1609.17 & &   0.986  &   0.993  &   0.990  & & -17.237 & -17.302 & -17.235 & -17.237 &   1.000 \\
    5 & 4137.18 & &  92.107  &  92.073  &  92.090  & & -18.328 & -18.323 & -18.327 & -18.326 &   1.000 \\
    6 & 4138.01 & &  26.788  &  26.766  &  26.777  & &   2.805 &   2.804 &   2.805 &   2.805 &   1.000 \\
    \bottomrule
    \end{tabular*}
\end{table*}

\subsection{Method Comparison}
In discussing the accuracy of the VG and LG(OI) approaches to origin
invariance, we build our analysis around comparisons between the VG and LG(OI)
methods to GIAOs for \smo, \apn, and \cmp\ simulated at the B3PW91 level of
theory with the aug-cc-pVTZ and aug-cc-pVQZ basis sets, specifically for the
experimentally relevant region of the VCD spectrum ($700-2000$ cm$^{-1}$).

\begin{figure*}
\centering
    \begin{subfigure}[t]{\textwidth}
        \stackunder[3pt]{\includegraphics[width=6.4in,height=2.5in]{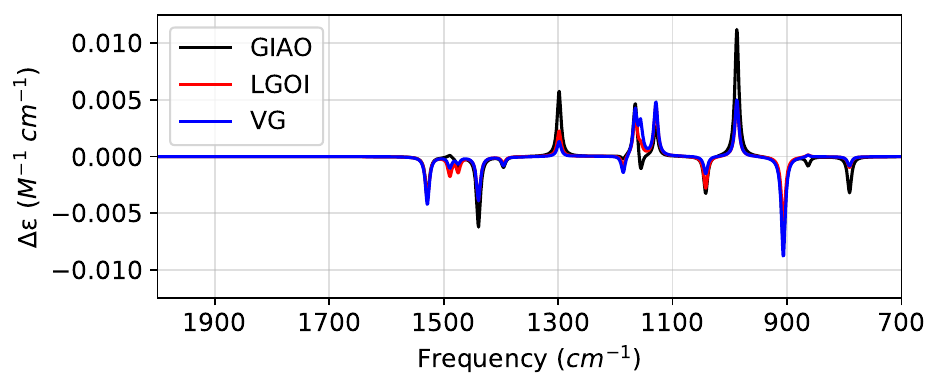}}{\ \ \ \ \ {\large (a)}}
    \end{subfigure} \\
    \begin{subfigure}[t]{\textwidth}
        \stackunder[3pt]{\includegraphics[width=6.4in,height=2.5in]{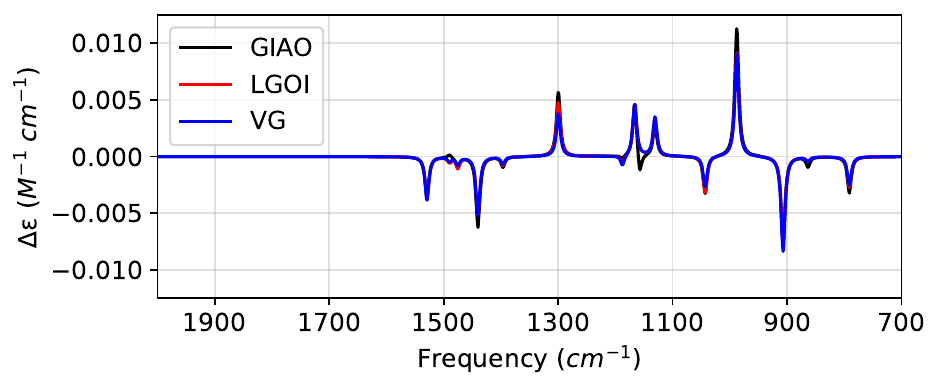}}{\ \ \ \ \ {\large (b)}}
    \end{subfigure}
\caption{VCD spectra of \smo\ computed at the (a) B3PW91/aug-cc-pVTZ
        and (b) B3PW91/aug-cc-pVQZ levels of theory.}
\label{fig:smo_B3PW91_spectra}
\end{figure*}

The VCD spectrum of \smo\ at the B3PW91/aug-cc-pVTZ and /aug-cc-pVQZ levels of
theory is presented in Fig.~\ref{fig:smo_B3PW91_spectra}, along with the
corresponding frequencies, rotational strengths, and rotational strength
differences in Table~\ref{tab:smo_b3pw91_data}.  We observe good agreement with
the VG and LG(OI) methods for the major peaks, specifically around 907 cm$^{-1}$
and 988 cm$^{-1}$, using the aug-cc-pVQZ basis set. In contrast, we note the
peak at 988 cm$^{-1}$ deviates significantly from the GIAO value using the
aug-cc-pVTZ basis.  Perhaps more notably, however, is the sign discrepancy at
approximately 1156 cm$^{-1}$ for which we observe differences between the GIAO
and VG approaches (denoted $\Delta R_i^{VG}$) nearly doubling the magnitude of
GIAO results using the aug-cc-pVTZ basis.  The difference between the GIAO and
LG(OI) methods ($\Delta R_i^{LGOI}$) for this mode show improvement over the VG
approach but still exhibits the wrong sign.  For the aug-cc-pVQZ basis set we
note significant improvements for both the VG and LG(OI) approaches for this
peak, though the signs still differ from that of the GIAO method.  A relatively
weak mode at approximately 1489 cm$^{-1}$ also displays sign discrepancies using
the aug-cc-pVTZ and aug-cc-pVQZ basis sets for the VG and LG(OI) approaches
compared to the GIAO approach.  In general, the number of sign discrepancies
between the GIAO method vs.\ the VG and LG(OI) approaches decreases from three
to two between the aug-cc-pVTZ and aug-cc-pVQZ basis sets.  Additionally, root
mean square (RMS) values of $\Delta R_i^{VG}$ and $\Delta R_i^{LGOI}$ show
improved accuracy with increasing basis set size for the LG(OI) method over the
VG method with values of 7.10 and 2.79 for the VG approach and 6.50 and 2.41 for
the LG(OI) approach using the aug-cc-pVTZ and aug-cc-pVQZ basis sets,
respectively.  Finally, we find limited correspondence between the DoS and the
accuracy of the LG(OI) method relative to the GIAO approach.  For example, mode
14 computed with the aug-cc-pVTZ basis displays a DoS of 0.299, even though the
deviation of this weak mode is only -0.725 while mode eight has a DoS of 0.971
with a $\Delta R_i^{LGOI}$ of -1.352.

\begin{table*}
\scriptsize
\caption{Frequencies (cm$^{-1}$), rotational strengths (10$^{-44}$ esu$^{2}$ cm$^{2}$),
        rotational strength errors relative to the use of GIAOs, and degrees of symmetry for \smo\ 
        using the B3PW91 functional with the aug-cc-pVTZ and aug-cc-pVQZ basis sets.}
\label{tab:smo_b3pw91_data}
\renewcommand{\arraystretch}{0.6}
\begin{tabular*}{\textwidth}{@{\extracolsep{\fill}}lrrrrrrrrrrrrrrr}
\toprule
& & \multicolumn{6}{c}{aug-cc-pVTZ} & \hspace{1pt} & \multicolumn{7}{c}{aug-cc-pVQZ} \\
\cmidrule{2-8} \cmidrule{10-16}
Mode & \multicolumn{1}{r}{Frequency} & \multicolumn{1}{r}{$R_i^{VG}$} & \multicolumn{1}{r}{$R_i^{LGOI}$} & \multicolumn{1}{r}{$R_i^{GIAO}$} & \multicolumn{1}{r}{$\Delta R_i^{VG}$} & \multicolumn{1}{r}{$\Delta R_i^{LGOI}$} & \multicolumn{1}{r}{DoS} & &
\multicolumn{1}{r}{Frequency} & \multicolumn{1}{r}{$R_i^{VG}$} & \multicolumn{1}{r}{$R_i^{LGOI}$} & \multicolumn{1}{r}{$R_i^{GIAO}$} & \multicolumn{1}{r}{$\Delta R_i^{VG}$} & \multicolumn{1}{r}{$\Delta R_i^{LGOI}$} & \multicolumn{1}{r}{DoS} \\
\midrule
\ \ \vdots & \vdots \ \ \ \ \ & \vdots \ \ \ \ & \vdots \ \ \ \ & \vdots \ \ \ \ & \vdots \ \ \ \ & \vdots \ \ \ \ & \vdots \ \ \ \ 
& & \vdots \ \ \ \ \ & \vdots \ \ \ \ & \vdots \ \ \ \ & \vdots \ \ \ \ & \vdots \ \ \ \ & \vdots \ \ \ \ & \vdots \ \ \ \ \\
  4 &  790.91 &  -3.059 &  -3.504 & -11.677 &  -8.618 & -8.174 & 0.742 & &  791.09 &  -9.254 &  -9.920 & -11.774 & -2.520 & -1.854 & 0.936 \\ 
  5 &  863.27 &   0.670 &   0.642 &  -2.607 &  -3.276 & -3.248 & 0.934 & &  863.41 &  -1.388 &  -1.376 &  -3.006 & -1.618 & -1.630 & 0.976 \\
  6 &  906.54 & -28.333 & -19.658 & -25.100 &   3.233 & -5.442 & 0.568 & &  906.99 & -26.694 & -24.378 & -24.503 &  2.191 & -0.125 & 0.801 \\
  7 &  987.74 &  14.746 &  14.271 &  33.035 &  18.290 & 18.765 & 0.785 & &  987.87 &  26.534 &  26.851 &  33.098 &  6.564 &  6.246 & 0.927 \\
  8 & 1042.28 &  -4.311 &  -7.887 &  -9.239 &  -4.928 & -1.352 & 0.971 & & 1043.08 &  -7.537 &  -8.948 &  -9.195 & -1.658 & -0.247 & 0.996 \\
  9 & 1129.29 &  12.170 &  11.440 &   6.753 &  -5.417 & -4.687 & 0.839 & & 1130.77 &   8.864 &   8.623 &   7.027 & -1.837 & -1.596 & 0.955 \\
 10 & 1155.86 &   6.672 &   1.830 &  -4.614 & -11.286 & -6.444 & 0.430 & & 1157.50 &   0.018 &   0.011 &  -5.109 & -5.127 & -5.120 & 0.568 \\
 11 & 1165.30 &   9.593 &   9.144 &  12.298 &   2.705 &  3.154 & 0.881 & & 1166.22 &  11.329 &  11.247 &  12.323 &  0.994 &  1.077 & 0.962 \\
 12 & 1186.26 &  -3.944 &  -2.354 &  -0.880 &   3.064 &  1.475 & 0.691 & & 1187.55 &  -2.241 &  -1.825 &  -0.764 &  1.477 &  1.061 & 0.847 \\
 13 & 1298.36 &   2.957 &   5.072 &  12.912 &   9.956 &  7.840 & 0.592 & & 1299.67 &   8.480 &  10.667 &  12.673 &  4.193 &  2.006 & 0.894 \\
 14 & 1395.51 &  -1.631 &  -1.200 &  -1.925 &  -0.294 & -0.725 & 0.299 & & 1396.48 &  -1.370 &  -1.590 &  -1.918 & -0.549 & -0.328 & 0.581 \\
 15 & 1439.14 &  -7.801 &  -7.723 & -12.476 &  -4.674 & -4.752 & 0.896 & & 1440.07 & -10.358 & -10.297 & -12.506 & -2.148 & -2.209 & 0.959 \\
 16 & 1474.71 &  -1.063 &  -2.478 &  -1.791 &  -0.729 &  0.686 & 0.386 & & 1475.31 &  -1.423 &  -1.942 &  -1.793 & -0.370 &  0.150 & 0.868 \\
 17 & 1489.18 &  -1.809 &  -3.128 &   0.449 &   2.258 &  3.578 & 0.569 & & 1489.79 &  -0.715 &  -0.870 &   0.527 &  1.242 &  1.397 & 0.888 \\
 18 & 1528.57 &  -7.999 &  -7.181 &  -6.836 &   1.162 &  0.344 & 0.741 & & 1529.27 &  -7.265 &  -7.087 &  -6.730 &  0.535 &  0.356 & 0.898 \\
\ \ \vdots & \vdots \ \ \ \ \ & \vdots \ \ \ \ & \vdots \ \ \ \ & \vdots \ \ \ \ & \vdots \ \ \ \ & \vdots \ \ \ \ & \vdots \ \ \ \ 
& & \vdots \ \ \ \ \ & \vdots \ \ \ \ & \vdots \ \ \ \ & \vdots \ \ \ \ & \vdots \ \ \ \ & \vdots \ \ \ \ & \vdots \ \ \ \ \\
\bottomrule
\end{tabular*}
\end{table*}

\begin{figure*}
\centering
    \begin{subfigure}[t]{\textwidth}
        \stackunder[3pt]{\includegraphics[width=6.4in,height=2.5in]{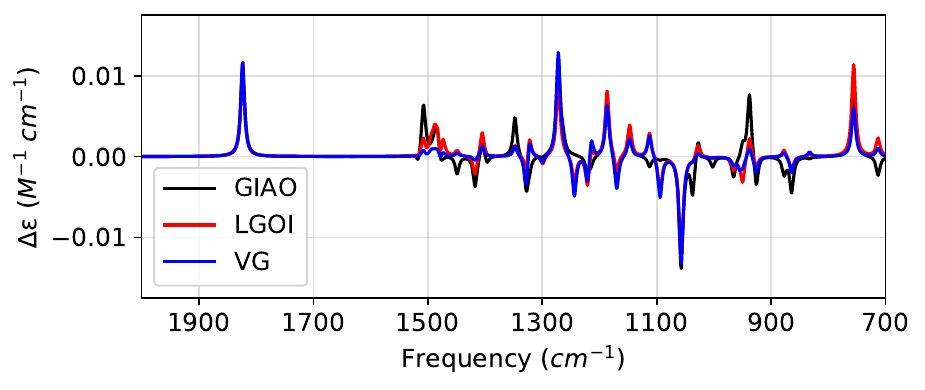}}{\ \ \ \ \ (a)}
    \end{subfigure} \\
    \begin{subfigure}[t]{\textwidth}
        \stackunder[3pt]{\includegraphics[width=6.4in,height=2.5in]{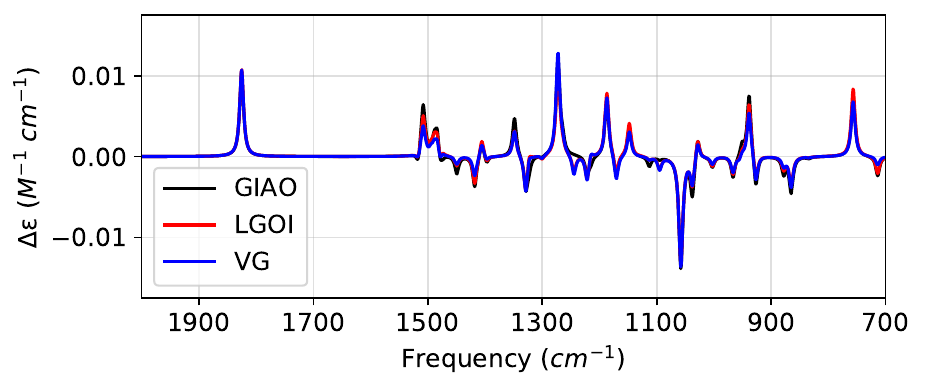}}{\ \ \ \ \ (b)}
    \end{subfigure}
\caption{VCD spectra of \cmp\ optimized and computed at the (a) B3PW91/aug-cc-pVTZ
        and (b) B3PW91/aug-cc-pVQZ levels of theory.}
\label{fig:cmp_B3PW91_spectra}
\end{figure*}

In Fig.~\ref{fig:cmp_B3PW91_spectra} and Table~\ref{tab:cmp_b3pw91_data} we
present the spectra, frequencies, rotational strengths, and rotational strength
differences for \cmp\ computed with the VG, LG(OI), and GIAO approaches using
the aug-cc-pVTZ and aug-cc-pVQZ basis sets.  As for \smo, we observe good
agreement for the major peaks, specifically the modes at approximately 756
cm$^{-1}$, 938 cm$^{-1}$, 1057 cm$^{-1}$, and 1272 cm$^{-1}$, computed with VG,
LG(OI), and GIAO approach using the aug-cc-pVQZ basis.  For the aug-cc-pVTZ
basis, however, there are some significant discrepancies in the intensities of
these major peaks.  For example, the GIAO method predicts the mode at 756
cm$^{-1}$ to have a rotational strength of $27.926 \times 10^{-44}$ esu$^2$ cm$^2$
while the VG and LG(OI) methods predict rotational strengths of $22.711 \times
10^{-44}$ esu$^2$ cm$^2$ and $43.469 \times 10^{-44}$ esu$^2$ cm$^2$,
respectively.  Though the VG approach predicts a reasonable rotational strength,
the LG(OI) method predicts almost two times the intensity of the GIAO method.
Similarly, the GIAO method predicts the mode at 938 cm$^{-1}$ to be $24.702
\times 10^{-44}$ esu$^2$ cm$^2$ while the VG and LG(OI) methods predict
rotational strengths of $4.159 \times 10^{-44}$ esu$^2$ cm$^2$ and $8.445 \times
10^{-44}$ esu$^2$ cm$^2$, respectively.  Of the four sign discrepancies between
the VG and LG(OI) methods and that of the GIAO method using the aug-cc-pVQZ
basis set, only the discrepancy at 1244.63 cm$^{-1}$ is noticeable in the
spectra.  This observation contrasts significantly from what we observe for the
aug-cc-pVTZ spectra where eight of the 11 sign discrepancies are easily
discernible in the spectra.  Regarding the RMS errors for the VG/LG(OI) methods,
we observe improvements going from the aug-cc-pVTZ to the aug-cc-pVQZ basis set
of 7.08/7.33 to 2.75/2.25, respectively.  Additionally, we find that the number
of modes where the LG(OI) approach outperforms the VG method increases from 22
to 29 for 41 modes in the experimentally relevant region.

\begin{table*}
\scriptsize
\caption{Frequencies (cm$^{-1}$), rotational strengths (10$^{-44}$ esu$^{2}$ cm$^{2}$),
        rotational strength errors, and degrees of symmetry for \cmp\ optimized and
        computed at the B3PW91/aug-cc-pVTZ and B3PW91/aug-cc-pVQZ levels of theory.}
\label{tab:cmp_b3pw91_data}
\renewcommand{\arraystretch}{0.6}
\begin{tabular*}{\textwidth}{@{\extracolsep{\fill}}lrrrrrrrrrrrrrrr}
\toprule
& & \multicolumn{6}{c}{aug-cc-pVTZ} & \hspace{1pt} & \multicolumn{7}{c}{aug-cc-pVQZ} \\
\cmidrule{2-8} \cmidrule{10-16}
Mode & \multicolumn{1}{r}{Frequency} & \multicolumn{1}{r}{$R_i^{VG}$} & \multicolumn{1}{r}{$R_i^{LGOI}$} & \multicolumn{1}{r}{$R_i^{GIAO}$} & \multicolumn{1}{r}{$\Delta R_i^{VG}$} & \multicolumn{1}{r}{$\Delta R_i^{LGOI}$} & \multicolumn{1}{r}{DoS} & &
\multicolumn{1}{r}{Frequency} & \multicolumn{1}{r}{$R_i^{VG}$} & \multicolumn{1}{r}{$R_i^{LGOI}$} & \multicolumn{1}{r}{$R_i^{GIAO}$} & \multicolumn{1}{r}{$\Delta R_i^{VG}$} & \multicolumn{1}{r}{$\Delta R_i^{LGOI}$} & \multicolumn{1}{r}{DoS} \\
\midrule
\ \ \vdots & \vdots \ \ \ \ \ & \vdots \ \ \ \ & \vdots \ \ \ \ & \vdots \ \ \ \ & \vdots \ \ \ \ & \vdots \ \ \ \ & \vdots \ \ \ \
& & \vdots \ \ \ \ \ & \vdots \ \ \ \ & \vdots \ \ \ \ & \vdots \ \ \ \ & \vdots \ \ \ \ & \vdots \ \ \ \ & \vdots \ \ \ \ \\
 19 &  713.09 &   4.017 &   9.047 &  -9.547 & -13.564 & -18.594 & 0.801 & &  713.88 &  -4.305 &  -8.673 &  -9.739 & -5.434 & -1.066 & 0.880 \\
 20 &  755.84 &  22.711 &  43.469 &  27.926 &   5.215 & -15.543 & 0.761 & &  756.69 &  26.072 &  32.145 &  28.167 &  2.095 & -3.978 & 0.947 \\
 21 &  832.84 &   2.024 &   0.975 &  -0.778 &  -2.801 &  -1.752 & 0.374 & &  833.56 &  -0.446 &  -0.409 &  -0.745 & -0.300 & -0.337 & 0.713 \\
 22 &  864.31 &  -6.861 &  -5.801 & -14.586 &  -7.725 &  -8.785 & 0.774 & &  865.04 & -12.512 & -11.179 & -14.710 & -2.198 & -3.531 & 0.901 \\
 23 &  877.28 &   2.115 &   3.100 &  -6.752 &  -8.867 &  -9.853 & 0.314 & &  878.01 &  -3.650 &  -4.959 &  -6.691 & -3.041 & -1.732 & 0.818 \\
 24 &  926.08 &  -6.255 &  -4.620 & -13.129 &  -6.874 &  -8.509 & 0.586 & &  926.78 & -10.640 & -10.138 & -13.110 & -2.471 & -2.973 & 0.808 \\
 25 &  937.92 &   4.159 &   8.445 &  24.702 &  20.542 &  16.257 & 0.658 & &  938.49 &  17.926 &  21.077 &  24.469 &  6.543 &  3.392 & 0.941 \\
 26 &  949.83 &  -3.172 &  -9.769 &   4.593 &   7.766 &  14.363 & 0.294 & &  950.74 &   1.342 &   1.969 &   4.463 &  3.121 &  2.493 & 0.886 \\
 27 &  955.16 &  -3.828 &  -0.866 &  -1.015 &   2.813 &  -0.149 & 0.751 & &  955.90 &  -1.540 &  -1.380 &  -0.434 &  1.106 &  0.946 & 0.738 \\
 28 &  965.48 &  -2.825 &  -4.610 &  -8.023 &  -5.199 &  -3.414 & 0.871 & &  966.52 &  -6.071 &  -6.836 &  -8.147 & -2.077 & -1.312 & 0.961 \\
 29 & 1002.05 &  -0.217 &  -0.622 &  -3.562 &  -3.346 &  -2.940 & 0.705 & & 1002.68 &  -2.632 &  -3.334 &  -3.662 & -1.030 & -0.328 & 0.948 \\
 30 & 1027.82 &   1.774 &   4.692 &   7.552 &   5.778 &   2.860 & 0.762 & & 1028.46 &   6.322 &   7.551 &   8.083 &  1.761 &  0.533 & 0.964 \\
 31 & 1037.20 &  -0.794 &  -1.040 & -13.066 & -12.272 & -12.026 & 0.913 & & 1037.72 &  -9.587 & -10.194 & -13.667 & -4.079 & -3.472 & 0.968 \\
 32 & 1057.23 & -36.066 & -34.285 & -37.577 &  -1.511 &  -3.292 & 0.972 & & 1057.77 & -37.036 & -36.260 & -37.524 & -0.489 & -1.265 & 0.990 \\
 33 & 1094.06 & -13.326 & -11.088 &  -0.691 &  12.635 &  10.397 & 0.847 & & 1094.89 &  -4.075 &  -3.746 &  -0.693 &  3.382 &  3.053 & 0.955 \\
 34 & 1112.46 &   7.436 &   8.170 &  -3.201 & -10.637 & -11.370 & 0.540 & & 1112.83 &  -0.885 &  -0.925 &  -3.136 & -2.251 & -2.212 & 0.870 \\
 35 & 1147.21 &   5.771 &   9.839 &   9.119 &   3.348 &  -0.720 & 0.297 & & 1147.88 &   7.821 &  10.452 &   9.213 &  1.391 & -1.240 & 0.800 \\
 36 & 1169.80 & -10.759 &  -5.684 &  -6.784 &   3.975 &  -1.100 & 0.761 & & 1170.54 &  -7.937 &  -6.246 &  -6.712 &  1.225 & -0.466 & 0.875 \\
 37 & 1186.55 &  16.010 &  20.168 &  17.366 &   1.357 &  -2.802 & 0.500 & & 1187.07 &  18.093 &  19.380 &  17.287 & -0.806 & -2.092 & 0.821 \\
 38 & 1214.38 &   6.890 &   3.490 &  -1.981 &  -8.871 &  -5.471 & 0.359 & & 1215.14 &   2.233 &   1.932 &  -1.811 & -4.044 & -3.744 & 0.590 \\
 39 & 1220.79 &  -9.098 &  -9.522 &  -5.220 &   3.878 &   4.302 & 0.694 & & 1221.45 &  -7.741 &  -7.411 &  -5.360 &  2.381 &  2.051 & 0.906 \\
 40 & 1243.82 & -11.935 &  -7.887 &   0.096 &  12.031 &   7.983 & 0.528 & & 1244.63 &  -5.653 &  -4.696 &   0.018 &  5.671 &  4.714 & 0.758 \\
 41 & 1263.86 &   1.719 &   0.534 &   3.571 &   1.853 &   3.038 & 0.293 & & 1264.55 &   1.793 &   1.654 &   3.410 &  1.617 &  1.757 & 0.306 \\
 42 & 1271.92 &  29.227 &  20.255 &  26.447 &  -2.780 &   6.192 & 0.903 & & 1272.50 &  29.145 &  24.975 &  26.963 & -2.182 &  1.988 & 0.958 \\
 43 & 1299.20 &  -2.277 &  -2.518 &  -0.619 &   1.658 &   1.899 & 0.489 & & 1300.08 &  -0.971 &  -1.114 &  -0.618 &  0.352 &  0.496 & 0.783 \\
 44 & 1322.48 &   6.613 &   6.347 &  -2.537 &  -9.151 &  -8.884 & 0.306 & & 1323.41 &   2.222 &   2.765 &  -2.725 & -4.947 & -5.490 & 0.611 \\
 45 & 1327.70 & -10.023 &  -5.612 &  -8.958 &   1.064 &  -3.346 & 0.302 & & 1328.56 & -10.570 & -10.290 &  -8.844 &  1.727 &  1.446 & 0.491 \\
 46 & 1347.61 &   2.913 &   2.930 &  10.687 &   7.774 &   7.757 & 0.713 & & 1348.47 &   6.979 &   7.104 &  10.535 &  3.556 &  3.432 & 0.878 \\
 47 & 1396.39 &  -0.067 &  -0.423 &  -1.971 &  -1.904 &  -1.548 & 0.576 & & 1397.25 &  -1.158 &  -1.733 &  -1.985 & -0.826 & -0.252 & 0.962 \\
 48 & 1404.93 &   2.648 &   6.570 &   4.319 &   1.671 &  -2.250 & 0.438 & & 1405.54 &   3.536 &   4.845 &   4.306 &  0.770 & -0.538 & 0.887 \\
 49 & 1417.15 &  -1.183 &  -5.165 &  -7.953 &  -6.771 &  -2.788 & 0.798 & & 1418.01 &  -5.122 &  -7.199 &  -7.909 & -2.788 & -0.710 & 0.975 \\
 50 & 1448.72 &   0.769 &   1.349 &  -4.306 &  -5.075 &  -5.655 & 0.811 & & 1449.25 &  -1.968 &  -2.347 &  -4.305 & -2.338 & -1.958 & 0.949 \\
 51 & 1473.35 &   1.094 &   4.240 &  -0.709 &  -1.803 &  -4.948 & 0.623 & & 1473.99 &   0.598 &   0.832 &  -0.734 & -1.332 & -1.566 & 0.943 \\
 52 & 1478.08 &  -1.352 &  -3.236 &  -2.982 &  -1.630 &   0.254 & 0.712 & & 1478.64 &  -2.060 &  -2.642 &  -3.038 & -0.978 & -0.396 & 0.946 \\
 53 & 1482.97 &   1.494 &   5.015 &   5.950 &   4.455 &   0.934 & 0.693 & & 1483.55 &   3.472 &   4.689 &   6.039 &  2.567 &  1.350 & 0.953 \\
 54 & 1487.94 &   0.937 &   4.700 &   3.805 &   2.868 &  -0.895 & 0.708 & & 1488.51 &   2.301 &   3.290 &   3.712 &  1.412 &  0.422 & 0.971 \\
 55 & 1493.57 &   1.303 &   3.267 &   0.843 &  -0.460 &  -2.424 & 0.725 & & 1494.13 &   1.403 &   1.826 &   0.778 & -0.625 & -1.048 & 0.946 \\
 56 & 1503.64 &  -0.266 &  -0.501 &   0.996 &   1.262 &   1.497 & 0.309 & & 1504.18 &   0.661 &   1.051 &   1.099 &  0.438 &  0.048 & 0.814 \\
 57 & 1507.64 &   1.467 &   4.198 &  11.975 &  10.508 &   7.776 & 0.804 & & 1508.19 &   6.932 &   9.125 &  11.934 &  5.002 &  2.809 & 0.970 \\
 58 & 1516.46 &  -0.140 &  -0.463 &  -2.711 &  -2.571 &  -2.248 & 0.804 & & 1516.89 &  -1.309 &  -1.767 &  -2.750 & -1.442 & -0.983 & 0.969 \\
 59 & 1823.50 &  18.677 &  18.196 &  16.589 &  -2.089 &  -1.607 & 0.997 & & 1825.31 &  17.015 &  17.070 &  16.587 & -0.428 & -0.482 & 0.999 \\
\ \ \vdots & \vdots \ \ \ \ \ & \vdots \ \ \ \ & \vdots \ \ \ \ & \vdots \ \ \ \ & \vdots \ \ \ \ & \vdots \ \ \ \ & \vdots \ \ \ \
& & \vdots \ \ \ \ \ & \vdots \ \ \ \ & \vdots \ \ \ \ & \vdots \ \ \ \ & \vdots \ \ \ \ & \vdots \ \ \ \ & \vdots \ \ \ \ \\
\bottomrule
\end{tabular*}
\end{table*}



\begin{figure*}
\centering
    \begin{subfigure}[t]{\textwidth}
        \stackunder[3pt]{\includegraphics[width=6.4in,height=2.5in]{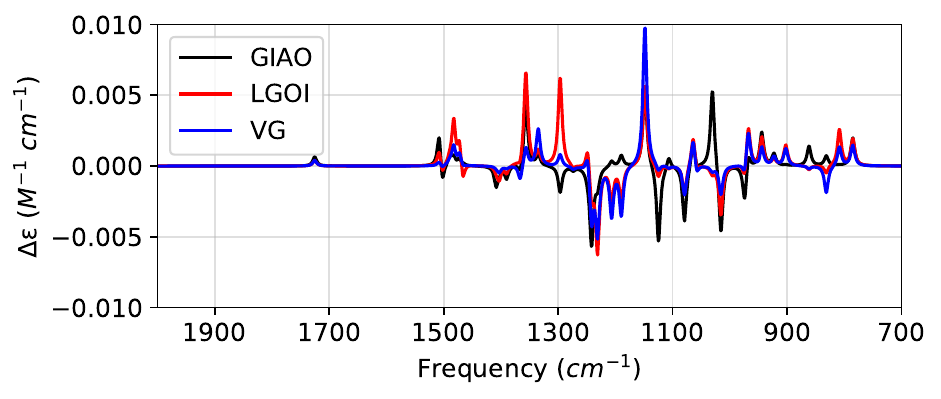}}{\ \ \ \ \ (a)}
    \end{subfigure} \\
    \begin{subfigure}[t]{\textwidth}
        \stackunder[3pt]{\includegraphics[width=6.4in,height=2.5in]{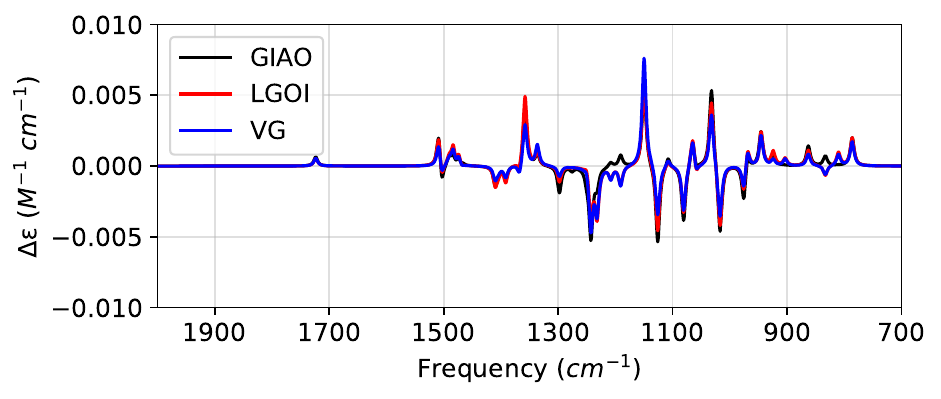}}{\ \ \ \ \ (b)}
    \end{subfigure}
\caption{VCD spectra of \apn\ optimized and computed at the (a) B3PW91/aug-cc-pVTZ
        and (b) B3PW91/aug-cc-pVQZ levels of theory.} 
\label{fig:apn_B3PW91_spectra}
\end{figure*}

For our final test molecule, \apn, we provide spectra, frequencies, rotational
strengths, and rotational strength differences in
Fig.~\ref{fig:apn_B3PW91_spectra} and Table~\ref{tab:apn_b3pw91_data}.  As was
the case for \smo\ and \cmp, we observe good agreement between the VG and LG(OI)
formulations with the GIAO approach for the major peaks (modes at approximately
1016 cm$^{-1}$, 1031 cm$^{-1}$, 1125 cm$^{-1}$, and 1149 cm$^{-1}$) using the
aug-cc-pVQZ basis set.  At the aug-cc-pVTZ level, the agreement is much worse
including a sign error for the mode at ca.\ 1031 cm$^{-1}$, overestimation of
the rotational strengths for the modes at 1297 and 1356 cm$^{-1}$, and
underestimation of the peak at 1125 cm$^{-1}$.  At the aug-cc-pVQZ level,
Table~\ref{tab:apn_b3pw91_data} indicates that there are a total of five sign
discrepancies between LG(OI) and GIAO, though only three of these are visible in
the spectra in Fig.~\ref{fig:apn_B3PW91_spectra} at approximately 832 cm$^{-1}$,
1190 cm$^{-1}$, and 1207 cm$^{-1}$.  The aug-cc-pVTZ basis set, on the other
hand, exhibits a total of ten sign discrepancies between the VG and LG(OI)
methods with those of the GIAO method for the 40 experimentally relevant modes.
As noted for the other test cases, there is a clear improvement in the RMS
errors for the VG/LG(OI) approaches when going from aug-cc-pVTZ to aug-cc-pVQZ
with values of 5.68/5.95 and 2.37/1.94, respectively.  Interestingly, these
improvements are not as substantial as those noted for \smo\ and \apn.  Unlike
both \smo\ and \apn, however, there is very little improvement in the number of
modes where the LG(OI) method provides an advantage over the VG method in terms
of having smaller absolute differences from the GIAO method.

\begin{table*}
\scriptsize
\caption{Frequencies (cm$^{-1}$), rotational strengths (10$^{-44}$ esu$^{2}$ cm$^{2}$),
        rotational strength errors, and degrees of symmetry for \apn\ optimized and 
        computed at the B3PW91/aug-cc-pVTZ and B3PW91/aug-cc-pVQZ levels of theory.}
\label{tab:apn_b3pw91_data}
\renewcommand{\arraystretch}{0.6}
\begin{tabular*}{\textwidth}{@{\extracolsep{\fill}}lrrrrrrrrrrrrrrr}
\toprule
& & \multicolumn{6}{c}{aug-cc-pVTZ} & \hspace{1pt} & \multicolumn{7}{c}{aug-cc-pVQZ} \\
\cmidrule{2-8} \cmidrule{10-16}
Mode & \multicolumn{1}{r}{Frequency} & \multicolumn{1}{r}{$R_i^{VG}$} & \multicolumn{1}{r}{$R_i^{LGOI}$} & \multicolumn{1}{r}{$R_i^{GIAO}$} & \multicolumn{1}{r}{$\Delta R_i^{VG}$} & \multicolumn{1}{r}{$\Delta R_i^{LGOI}$} & \multicolumn{1}{r}{DoS} & &
\multicolumn{1}{r}{Frequency} & \multicolumn{1}{r}{$R_i^{VG}$} & \multicolumn{1}{r}{$R_i^{LGOI}$} & \multicolumn{1}{r}{$R_i^{GIAO}$} & \multicolumn{1}{r}{$\Delta R_i^{VG}$} & \multicolumn{1}{r}{$\Delta R_i^{LGOI}$} & \multicolumn{1}{r}{DoS} \\
\midrule
\ \ \vdots & \vdots \ \ \ \ \ & \vdots \ \ \ \ & \vdots \ \ \ \ & \vdots \ \ \ \ & \vdots \ \ \ \ & \vdots \ \ \ \ & \vdots \ \ \ \
& & \vdots \ \ \ \ \ & \vdots \ \ \ \ & \vdots \ \ \ \ & \vdots \ \ \ \ & \vdots \ \ \ \ & \vdots \ \ \ \ & \vdots \ \ \ \ \\
 17 &  785.06 &   5.387 &   6.932 &   7.302 &   1.915 &   0.370 & 0.863 & &  786.24 &   6.277 &   7.325 &   7.374 &  1.096 &   0.049 &   0.966 \\
 18 &  808.56 &   4.961 &   9.199 &   0.133 &  -4.828 &  -9.067 & 0.859 & &  810.28 &   2.825 &   3.404 &   0.110 & -2.715 &  -3.294 &   0.967 \\
 19 &  831.47 &  -6.768 &  -2.031 &   2.433 &   9.201 &   4.464 & 0.567 & &  833.12 &  -2.271 &  -2.514 &   2.383 &  4.654 &   4.897 &   0.425 \\
 20 &  861.71 &  -0.652 &  -0.959 &   4.666 &   5.319 &   5.625 & 0.294 & &  862.89 &   2.789 &   3.754 &   4.714 &  1.925 &   0.960 &   0.819 \\
 21 &  902.18 &   3.879 &   4.695 &   0.189 &  -3.690 &  -4.506 & 0.615 & &  903.34 &   1.564 &   1.709 &   0.069 & -1.495 &  -1.640 &   0.878 \\
 22 &  922.99 &   1.961 &   1.337 &   2.624 &   0.663 &   1.287 & 0.332 & &  924.17 &   1.320 &   3.152 &   2.607 &  1.287 &  -0.545 &   0.451 \\
 23 &  944.25 &   4.282 &   6.452 &   7.363 &   3.080 &   0.911 & 0.618 & &  945.56 &   6.840 &   7.758 &   7.480 &  0.640 &  -0.279 &   0.919 \\
 24 &  948.37 &  -0.713 &  -0.702 &  -0.073 &   0.640 &   0.629 & 0.441 & &  950.16 &  -0.523 &  -0.980 &  -0.044 &  0.480 &   0.936 &   0.634 \\
 25 &  967.68 &   7.668 &   8.880 &   3.523 &  -4.144 &  -5.357 & 0.615 & &  969.01 &   3.902 &   4.450 &   3.325 & -0.577 &  -1.125 &   0.919 \\
 26 &  974.19 &  -2.705 &  -3.744 &  -7.835 &  -5.129 &  -4.091 & 0.454 & &  975.60 &  -5.274 &  -6.089 &  -7.766 & -2.492 &  -1.677 &   0.817 \\
 27 & 1015.61 &  -5.736 &  -9.877 & -14.187 &  -8.451 &  -4.310 & 0.830 & & 1017.11 & -10.802 & -12.804 & -14.111 & -3.310 &  -1.307 &   0.957 \\
 28 & 1030.48 &  -0.910 &  -1.338 &  15.863 &  16.773 &  17.201 & 0.389 & & 1032.10 &  10.852 &  13.395 &  15.923 &  5.071 &   2.529 &   0.855 \\
 29 & 1057.73 &  -2.131 &  -1.417 &  -1.062 &   1.069 &   0.355 & 0.297 & & 1058.81 &  -1.833 &  -2.169 &  -1.147 &  0.686 &   1.022 &   0.632 \\
 30 & 1064.17 &   5.550 &   5.937 &   5.256 &  -0.294 &  -0.680 & 0.587 & & 1064.98 &   5.429 &   5.966 &   5.284 & -0.145 &  -0.683 &   0.846 \\
 31 & 1079.12 &  -5.787 &  -5.394 & -10.753 &  -4.966 &  -5.360 & 0.815 & & 1080.75 &  -8.707 &  -9.100 & -10.604 & -1.897 &  -1.504 &   0.905 \\
 32 & 1106.72 &  -0.446 &  -0.451 &   2.109 &   2.555 &   2.561 & 0.876 & & 1107.94 &   1.314 &   1.280 &   2.025 &  0.711 &   0.745 &   0.968 \\
 33 & 1124.68 &  -1.324 &  -2.226 & -14.113 & -12.788 & -11.887 & 0.379 & & 1126.01 &  -9.348 & -12.135 & -14.152 & -4.804 &  -2.017 &   0.843 \\
 34 & 1148.43 &  25.002 &  14.547 &  14.310 & -10.692 &  -0.237 & 0.699 & & 1149.83 &  19.483 &  15.312 &  14.316 & -5.167 &  -0.997 &   0.853 \\
 35 & 1189.63 &  -8.326 &  -5.813 &   1.860 &  10.185 &   7.672 & 0.436 & & 1190.84 &  -3.360 &  -3.193 &   1.887 &  5.247 &   5.080 &   0.687 \\
 36 & 1206.59 &  -8.243 &  -5.765 &   0.991 &   9.234 &   6.756 & 0.445 & & 1208.05 &  -1.985 &  -1.859 &   0.757 &  2.743 &   2.616 &   0.790 \\
 37 & 1231.16 & -10.927 & -14.164 &  -2.921 &   8.006 &  11.242 & 0.555 & & 1232.12 &  -7.472 &  -8.191 &  -2.975 &  4.497 &   5.216 &   0.819 \\
 38 & 1241.65 &  -9.450 &  -5.594 & -12.758 &  -3.308 &  -7.164 & 0.578 & & 1242.73 & -10.393 &  -8.130 & -11.571 & -1.178 &  -3.441 &   0.762 \\
 39 & 1248.39 &   3.703 &   4.179 &  -0.674 &  -4.377 &  -4.853 & 0.535 & & 1249.96 &   1.179 &   1.290 &  -1.464 & -2.642 &  -2.753 &   0.802 \\
 40 & 1273.75 &  -0.450 &  -0.599 &  -0.532 &  -0.082 &   0.067 & 0.297 & & 1275.81 &  -0.046 &  -0.086 &  -0.635 & -0.589 &  -0.549 &   0.518 \\
 41 & 1296.51 &   1.861 &  13.983 &  -4.142 &  -6.003 & -18.125 & 0.373 & & 1297.78 &  -1.632 &  -2.512 &  -4.183 & -2.551 &  -1.671 &   0.932 \\
 42 & 1334.78 &   5.683 &   2.030 &   1.697 &  -3.986 &  -0.333 & 0.506 & & 1336.36 &   3.205 &   1.983 &   1.816 & -1.389 &  -0.167 &   0.669 \\
 43 & 1356.47 &   2.959 &  14.247 &   9.476 &   6.517 &  -4.770 & 0.293 & & 1357.86 &   6.441 &  10.710 &   9.291 &  2.850 &  -1.419 &   0.870 \\
 44 & 1366.43 &  -2.321 &  -1.163 &  -1.709 &   0.613 &  -0.546 & 0.443 & & 1367.66 &  -1.765 &  -1.378 &  -1.682 &  0.084 &  -0.304 &   0.642 \\
 45 & 1390.07 &  -0.347 &  -1.178 &  -1.865 &  -1.518 &  -0.686 & 0.420 & & 1391.64 &  -1.661 &  -2.397 &  -1.847 & -0.186 &   0.549 &   0.906 \\
 46 & 1403.13 &  -0.926 &  -1.959 &  -0.533 &   0.393 &   1.426 & 0.295 & & 1404.48 &  -0.572 &  -0.845 &  -0.605 & -0.033 &   0.240 &   0.800 \\
 47 & 1408.15 &  -0.147 &  -0.692 &  -2.854 &  -2.706 &  -2.162 & 0.663 & & 1409.85 &  -1.917 &  -2.750 &  -2.750 & -0.833 &   0.000 &   0.961 \\
 48 & 1466.70 &  -0.546 &  -3.047 &   0.514 &   1.060 &   3.561 & 0.584 & & 1467.56 &  -0.035 &  -0.052 &   0.567 &  0.603 &   0.619 &   0.956 \\
 49 & 1468.29 &   0.143 &   0.331 &  -0.522 &  -0.665 &  -0.853 & 0.940 & & 1469.19 &  -0.363 &  -0.454 &  -0.557 & -0.194 &  -0.104 &   0.988 \\
 50 & 1472.63 &   1.596 &   3.486 &   1.035 &  -0.561 &  -2.452 & 0.797 & & 1473.83 &   1.263 &   1.565 &   1.039 & -0.224 &  -0.526 &   0.953 \\
 51 & 1478.63 &  -0.283 &  -0.685 &  -0.187 &   0.095 &   0.498 & 0.869 & & 1479.67 &  -0.502 &  -0.629 &  -0.226 &  0.276 &   0.403 &   0.974 \\
 52 & 1482.39 &   2.783 &   6.487 &   1.209 &  -1.573 &  -5.277 & 0.851 & & 1483.32 &   2.226 &   2.808 &   1.254 & -0.972 &  -1.554 &   0.969 \\
 53 & 1489.30 &   0.513 &   0.687 &   1.217 &   0.704 &   0.530 & 0.422 & & 1490.30 &   0.881 &   1.083 &   1.199 &  0.318 &   0.116 &   0.817 \\
 54 & 1502.58 &  -0.406 &  -1.677 &  -3.370 &  -2.964 &  -1.693 & 0.695 & & 1503.52 &  -1.837 &  -2.559 &  -3.346 & -1.509 &  -0.787 &   0.963 \\
 55 & 1507.78 &   0.558 &   2.250 &   4.942 &   4.384 &   2.692 & 0.807 & & 1508.66 &   3.188 &   4.391 &   4.891 &  1.703 &   0.500 &   0.977 \\
 56 & 1725.43 &   0.677 &   0.650 &   1.117 &   0.440 &   0.468 & 0.719 & & 1723.60 &   0.918 &   0.882 &   1.089 &  0.170 &   0.207 &   0.870 \\
\ \ \vdots & \vdots \ \ \ \ \ & \vdots \ \ \ \ & \vdots \ \ \ \ & \vdots \ \ \ \ & \vdots \ \ \ \ & \vdots \ \ \ \ & \vdots \ \ \ \
& & \vdots \ \ \ \ \ & \vdots \ \ \ \ & \vdots \ \ \ \ & \vdots \ \ \ \ & \vdots \ \ \ \ & \vdots \ \ \ \ & \vdots \ \ \ \ \\
\bottomrule
\end{tabular*}
\end{table*}

%
%
%
%

%% file: conclusion.tex
\section{Discussion and Conclusion}

In this work, we have extended the LG(OI) approach involving the electric
dipole/magnetic dipole polarizability tensor from OR and ECD to VCD.  This
requires the reformulation of the electric dipole transition moment (APT) in the
VG as well as the diagonalization of a mixed LG/VG electric-dipole vibrational
polarizability tensor, which appears in the origin-dependent contributions to
the total LG rotational strength.  We have shown that both the VG and LG(OI)
approaches are origin invariant for VCD, as expected, and compared these results
to the GIAO approach.  Basis set convergence analyses demonstrate that the VG,
LG(OI), and GIAO methods converge to the same rotational strengths as the basis
set approaches completeness. 

For the test cases considered here [\smo, \cmp, and \apn], when deployed with
basis sets of quadruple-zeta quality, the LG(OI) and VG approaches yield VCD
spectra that compare well to those obtained using GIAOs, with relatively minor
discrepancies for all strong- and moderate-intensity peaks.  The advantage of
the LG(OI) approach over GIAO to achieve origin invariance is principally in the
ease of implementation, with the former requiring only the extension of existing
APT codes to use integrals over the linear-momentum operator.  This aspect may
be particularly advantageous for emerging VCD implementations using methods
including dynamic electron correlation effects.\cite{Shumberger2024, Shumberger2025,
Shumberger2026} In addition, the expected advantage of LG(OI) over the VG
formulation is that the LG-based rotational strength should converge more rapidly
with respect to basis-set completeness. Unfortunately, this is not the case for
the representative systems examined here, and we observe no substantial
improvement of the LG(OI) rotational strengths over their VG counterparts as
compared to GIAO results.


As an aside we feel that the LG(OI) method --- more specifically, its
underlying interpretation involving the rotations of the molecule along a 
principal axis --- provides interesting insight into the ability of a given
basis to provide a balanced space on which the different operators, i.e. the
position and momentum operators, may act.  By considering the SVD of
$[\vec{P}_i]^{LG}$ and $[\vec{P}_i]^{VG}$ separately where
\begin{align}
    [\vec{P}_i]^{LG} = \mathbf{U_i} [\vec{P}_i']^{LG} \mathbf{X_i^\dagger}
\end{align}
and
\begin{align}
    [\vec{P}_i]^{VG} = \mathbf{V_i} [\vec{P}_i']^{VG} \mathbf{Y_i^\dagger}
\end{align}
and their outer product becomes
\begin{align} \label{SVD2}
    \mathbf{[D_i']^{LG/VG}} = [\vec{P}_i']^{LG} \otimes [\vec{P}_i']^{VG}.
\end{align}
Eq.~\eqref{SVD2} can be shown to have a formal equivalence to the diagonal
quantity in Eq.~\eqref{SVD1} by noting that $\mathbf{X_i}$ and $\mathbf{Y_i} =
\mathbf{1}$, given the dimension of $[\vec{P}_i']^{LG}$ and $[\vec{P}_i']^{VG}$.
It becomes clear that $\mathbf{[D_i']^{LG/VG}}$ is a diagonal matrix with only
one nonzero element. As noted previously, the transformation matrices
$\mathbf{U_i}$ and $\mathbf{V_i}$ are interpreted as rotations which reorient
the molecule along the principal axis of the mixed LG/VG dipole-strength tensor
where $\mathbf{U_i}$ and $\mathbf{V_i}$ converge in the limit of a complete
basis set. From this alternative perspective, however, $\mathbf{U_i}$ and
$\mathbf{V_i}$ are distinct rotations of the molecule such that the length gauge
and velocity gauge electric dipole transition moments are now oriented along the
same axis leading to the idea that one can obtain a sort of "geometric error"
(attributed to inequalities in the description of different operators by a given
basis set and choice of wave function) from the difference between the
transformation matrices. As a result of the ambiguity in the phase and ordering
of the singular values and singular vectors, interpretation of the
transformation matrices as purely rotations is not straightforward. Analysis of
the determinant of the singular vectors results in $+1$ or $-1$ indicating
either a pure rotation or a rotation and reflection, respectively.  Assuming
that a singular vector includes a reflection, the reorientation has effectively
created the enantiomer.  Due to the fact that $\mathbf{[D_i']^{LG/VG}}$ has a
nullity of two, one can adjust the phase and ordering of the null-space singular
vectors at will to void any reflection character. The difference in these pure
rotations can be used to quantify geometrically differences in the basis-set
representation of the electric-dipole operator in the length and velocity
gauges.  In conclusion, not only can one obtain an origin-invariant formulation
of the electric-dipole/magnetic-dipole polarizability, but one may also consider
the transformation matrices from related operators as useful metrics in
evaluating the robustness of different basis sets.